\pdfoutput=1

\documentclass[12pt,a4paper]{article}

\usepackage{ifthen} 
\newboolean{pdflatex}
\setboolean{pdflatex}{true} 

\newboolean{articletitles}
\setboolean{articletitles}{true} 

\newboolean{uprightparticles}
\setboolean{uprightparticles}{false} 

\newboolean{inbibliography}
\setboolean{inbibliography}{false} 

\def\paperauthors{LHCb collaboration} 
\def\paperasciititle{Search for lepton-flavour-violating decays of Higgs-like bosons} 
\def\papertitle{Search for lepton-flavour-violating decays of Higgs-like bosons} 
\def\paperkeywords{{High Energy Physics}, {LHCb}, {Higgs}, {Muon}, {Tau}, {LFV}} 
\def\papercopyright{\the\year\ CERN for the benefit of the LHCb collaboration} 
\def\paperlicence{CC-BY-4.0 licence}
\def\paperlicenceurl{https://creativecommons.org/licenses/by/4.0/}

\usepackage[top=1in, bottom=1.25in, left=1in, right=1in]{geometry}

\usepackage{microtype}
\usepackage{lineno}  
\usepackage{xspace} 
\usepackage{caption} 

\usepackage{graphicx}  
\usepackage{color}
\usepackage{colortbl}
\graphicspath{{./figs/}{./figs/pdf/}} 
\DeclareGraphicsExtensions{.pdf,.PDF,png,.PNG}

\usepackage{amsmath} 
\usepackage{amssymb}
\usepackage{amsfonts}
\usepackage{upgreek} 

\newcommand*\patchAmsMathEnvironmentForLineno[1]{%
\expandafter\let\csname old#1\expandafter\endcsname\csname #1\endcsname
\expandafter\let\csname oldend#1\expandafter\endcsname\csname
end#1\endcsname
 \renewenvironment{#1}%
   {\linenomath\csname old#1\endcsname}%
   {\csname oldend#1\endcsname\endlinenomath}%
}
\newcommand*\patchBothAmsMathEnvironmentsForLineno[1]{%
  \patchAmsMathEnvironmentForLineno{#1}%
  \patchAmsMathEnvironmentForLineno{#1*}%
}
\AtBeginDocument{%
\patchBothAmsMathEnvironmentsForLineno{equation}%
\patchBothAmsMathEnvironmentsForLineno{align}%
\patchBothAmsMathEnvironmentsForLineno{flalign}%
\patchBothAmsMathEnvironmentsForLineno{alignat}%
\patchBothAmsMathEnvironmentsForLineno{gather}%
\patchBothAmsMathEnvironmentsForLineno{multline}%
\patchBothAmsMathEnvironmentsForLineno{eqnarray}%
}

\usepackage{hyperxmp}

\usepackage[pdftex,
            pdfauthor={\paperauthors},
            pdftitle={\paperasciititle},
            pdfkeywords={\paperkeywords},
            pdfcopyright={Copyright (C) \papercopyright},
            pdflicenseurl={\paperlicenceurl}]{hyperref}

\usepackage[all]{hypcap} 

\usepackage{xspace} 
\usepackage{upgreek}

\def\lhcb {\mbox{LHCb}\xspace}
\def\atlas  {\mbox{ATLAS}\xspace}
\def\cms    {\mbox{CMS}\xspace}

\def\babar  {\mbox{BaBar}\xspace}
\def\belle  {\mbox{Belle}\xspace}

\def\opal   {\mbox{OPAL}\xspace}

\def\lep    {\mbox{LEP}\xspace}

\def\presh  {PS\xspace}
\def\ecal   {ECAL\xspace}
\def\hcal   {HCAL\xspace}
\def\MagUp {\mbox{\em Mag\kern -0.05em Up}\xspace}

\ifthenelse{\boolean{uprightparticles}}%
{

 \def\Pmu         {\ensuremath{\upmu}\xspace}                 
 \def\Pnu         {\ensuremath{\upnu}\xspace}                 
                  
 \def\Ppi         {\ensuremath{\uppi}\xspace}

 \def\Ptau        {\ensuremath{\uptau}\xspace}

 \def\PDelta      {\ensuremath{\Delta}\xspace}                 
 \def\PXi      {\ensuremath{\Xi}\xspace}                 
 \def\PLambda      {\ensuremath{\Lambda}\xspace}                 
 \def\PSigma      {\ensuremath{\Sigma}\xspace}                 
 \def\POmega      {\ensuremath{\Omega}\xspace}                 
 \def\PUpsilon      {\ensuremath{\Upsilon}\xspace}                 
 
 \def\PB      {\ensuremath{\mathrm{B}}\xspace}                 
                  
 \def\PD      {\ensuremath{\mathrm{D}}\xspace}

 \def\PK      {\ensuremath{\mathrm{K}}\xspace}

 \def\PZ      {\ensuremath{\mathrm{Z}}\xspace}                 
                  
 \def\Pb      {\ensuremath{\mathrm{b}}\xspace}                 
 \def\Pc      {\ensuremath{\mathrm{c}}\xspace}                 
                  
 \def\Pe      {\ensuremath{\mathrm{e}}\xspace}

 \def\Pi      {\ensuremath{\mathrm{i}}\xspace}

 \def\Pt      {\ensuremath{\mathrm{t}}\xspace}

}
{

 \def\Pmu         {\ensuremath{\mu}\xspace}                 
 \def\Pnu         {\ensuremath{\nu}\xspace}                 
                  
 \def\Ppi         {\ensuremath{\pi}\xspace}

 \def\Ptau        {\ensuremath{\tau}\xspace}

 \mathchardef\PDelta="7101
 \mathchardef\PXi="7104
 \mathchardef\PLambda="7103
 \mathchardef\PSigma="7106
 \mathchardef\POmega="710A
 \mathchardef\PUpsilon="7107
                  
 \def\PB      {\ensuremath{B}\xspace}                 
                  
 \def\PD      {\ensuremath{D}\xspace}

 \def\PK      {\ensuremath{K}\xspace}

 \def\PZ      {\ensuremath{Z}\xspace}                 
                  
 \def\Pb      {\ensuremath{b}\xspace}                 
 \def\Pc      {\ensuremath{c}\xspace}                 
                  
 \def\Pe      {\ensuremath{e}\xspace}

 \def\Pi      {\ensuremath{i}\xspace}

 \def\Pt      {\ensuremath{t}\xspace}

}

\makeatletter
\ifcase \@ptsize \relax
  \newcommand{\miniscule}{\@setfontsize\miniscule{4}{5}}
\or
  \newcommand{\miniscule}{\@setfontsize\miniscule{5}{6}}
\or
  \newcommand{\miniscule}{\@setfontsize\miniscule{5}{6}}
\fi
\makeatother

\DeclareRobustCommand{\optbar}[1]{\shortstack{{\miniscule (\rule[.5ex]{1.25em}{.18mm})}
  \\ [-.7ex] $#1$}}

\def\electron   {{\ensuremath{\Pe}}\xspace}
\def\en         {{\ensuremath{\Pe^-}}\xspace}   

\def\epm        {{\ensuremath{\Pe^\pm}}\xspace} 
 
\def\epem       {{\ensuremath{\Pe^+\Pe^-}}\xspace}

\def\muon       {{\ensuremath{\Pmu}}\xspace}

\def\mun        {{\ensuremath{\Pmu^-}}\xspace} 
\def\mupm       {{\ensuremath{\Pmu^\pm}}\xspace} 
\def\mump       {{\ensuremath{\Pmu^\mp}}\xspace} 
\def\mumu       {{\ensuremath{\Pmu^+\Pmu^-}}\xspace}

\def\tauon      {{\ensuremath{\Ptau}}\xspace}

\def\taum       {{\ensuremath{\Ptau^-}}\xspace}
\def\taupm      {{\ensuremath{\Ptau^\pm}}\xspace}
\def\taump      {{\ensuremath{\Ptau^\mp}}\xspace}
\def\tautau     {{\ensuremath{\Ptau^+\Ptau^-}}\xspace}

\def\neu        {{\ensuremath{\Pnu}}\xspace}
\def\neub       {{\ensuremath{\overline{\Pnu}}}\xspace}

\def\neueb      {{\ensuremath{\neub_e}}\xspace}

\def\neumb      {{\ensuremath{\neub_\mu}}\xspace}
\def\neut       {{\ensuremath{\neu_\tau}}\xspace}

\def\Z      {{\ensuremath{\PZ}}\xspace}

\def\cquark    {{\ensuremath{\Pc}}\xspace}

\def\bquark    {{\ensuremath{\Pb}}\xspace}
\def\bquarkbar {{\ensuremath{\overline \bquark}}\xspace}
\def\bbbar     {{\ensuremath{\bquark\bquarkbar}}\xspace}
\def\tquark    {{\ensuremath{\Pt}}\xspace}
\def\tquarkbar {{\ensuremath{\overline \tquark}}\xspace}
\def\ttbar     {{\ensuremath{\tquark\tquarkbar}}\xspace}

\def\pion   {{\ensuremath{\Ppi}}\xspace}
\def\piz    {{\ensuremath{\pion^0}}\xspace}

\def\pip    {{\ensuremath{\pion^+}}\xspace}
\def\pim    {{\ensuremath{\pion^-}}\xspace}

  \def\Kbar    {{\kern 0.2em\overline{\kern -0.2em \PK}{}}\xspace}

\def\KorKbar    {\kern 0.18em\optbar{\kern -0.18em K}{}\xspace}

  \def\Dbar    {{\kern 0.2em\overline{\kern -0.2em \PD}{}}\xspace}

\def\DorDbar    {\kern 0.18em\optbar{\kern -0.18em D}{}\xspace}

\def\Bbar    {{\ensuremath{\kern 0.18em\overline{\kern -0.18em \PB}{}}}\xspace}

\def\BorBbar    {\kern 0.18em\optbar{\kern -0.18em B}{}\xspace}

  \def\Y#1S{\ensuremath{\PUpsilon{(#1S)}}\xspace}

\def\Lbar        {{\ensuremath{\kern 0.1em\overline{\kern -0.1em\PLambda}}}\xspace}
\def\LorLbar    {\kern 0.18em\optbar{\kern -0.18em \PLambda}{}\xspace}

\def\BF         {{\ensuremath{\mathcal{B}}}\xspace}

\def\BR         {\BF}
\newcommand{\decay}[2]{\ensuremath{#1\!\to #2}\xspace}         

\def\to                 {\ensuremath{\rightarrow}\xspace}

\def\AT#1     {\ensuremath{A_{\mathrm{T}}^{#1}}\xspace}           

\def\C#1      {\ensuremath{\mathcal{C}_{#1}}\xspace}                       
\def\Cp#1     {\ensuremath{\mathcal{C}_{#1}^{'}}\xspace}                    
\def\Ceff#1   {\ensuremath{\mathcal{C}_{#1}^{\mathrm{(eff)}}}\xspace}        
\def\Cpeff#1  {\ensuremath{\mathcal{C}_{#1}^{'\mathrm{(eff)}}}\xspace}       
\def\Ope#1    {\ensuremath{\mathcal{O}_{#1}}\xspace}                       
\def\Opep#1   {\ensuremath{\mathcal{O}_{#1}^{'}}\xspace}                    

\newcommand{\tev}{\ifthenelse{\boolean{inbibliography}}{\ensuremath{~T\kern -0.05em eV}}{\ensuremath{\mathrm{\,Te\kern -0.1em V}}}\xspace}
\newcommand{\gev}{\ensuremath{\mathrm{\,Ge\kern -0.1em V}}\xspace}
\newcommand{\mev}{\ensuremath{\mathrm{\,Me\kern -0.1em V}}\xspace}
\newcommand{\kev}{\ensuremath{\mathrm{\,ke\kern -0.1em V}}\xspace}
\newcommand{\ev}{\ensuremath{\mathrm{\,e\kern -0.1em V}}\xspace}
\newcommand{\gevc}{\ensuremath{{\mathrm{\,Ge\kern -0.1em V\!/}c}}\xspace}
\newcommand{\mevc}{\ensuremath{{\mathrm{\,Me\kern -0.1em V\!/}c}}\xspace}
\newcommand{\gevcc}{\ensuremath{{\mathrm{\,Ge\kern -0.1em V\!/}c^2}}\xspace}
\newcommand{\gevgevcccc}{\ensuremath{{\mathrm{\,Ge\kern -0.1em V^2\!/}c^4}}\xspace}
\newcommand{\mevcc}{\ensuremath{{\mathrm{\,Me\kern -0.1em V\!/}c^2}}\xspace}

\def\mum  {\ensuremath{{\,\upmu\mathrm{m}}}\xspace}

\def\pb {\ensuremath{\mathrm{ \,pb}}\xspace}
\def\invpb {\ensuremath{\mbox{\,pb}^{-1}}\xspace}

\def\invfb   {\ensuremath{\mbox{\,fb}^{-1}}\xspace}

\def\fs   {\ensuremath{\mathrm{ \,fs}}\xspace}

\def\gsim{{~\raise.15em\hbox{$>$}\kern-.85em
          \lower.35em\hbox{$\sim$}~}\xspace}
\def\lsim{{~\raise.15em\hbox{$<$}\kern-.85em
          \lower.35em\hbox{$\sim$}~}\xspace}

\def\sqs   {\ensuremath{\protect\sqrt{s}}\xspace}

\def\pt         {\ensuremath{p_{\mathrm{ T}}}\xspace}
\def\ptot       {\ensuremath{p}\xspace}

\def\rad{\ensuremath{\mathrm{ \,rad}}\xspace}

\newcommand{\lum} {\ensuremath{\mathcal{L}}\xspace}

\def\evtgen     {\mbox{\textsc{EvtGen}}\xspace}

\def\geant      {\mbox{\textsc{Geant4}}\xspace}

\def\photos     {\mbox{\textsc{Photos}}\xspace}

\def\pythia     {\mbox{\textsc{Pythia}}\xspace}

\def\tell1  {TELL1\xspace}
\def\ukl1   {UKL1\xspace}

\newcommand{\eg}{\mbox{\itshape e.g.}\xspace}
\newcommand{\ie}{\mbox{\itshape i.e.}\xspace}

\usepackage{cite} 
\usepackage{mciteplus}

\usepackage{mathtools}
\usepackage{booktabs,siunitx}
\usepackage{float} 
\usepackage{subfigure}

\hypersetup{
  colorlinks=true,
  citecolor=blue,
  linkcolor=red,
}

\usepackage{cleveref}
\crefname{section}{Sect.}{Sects.}
\crefname{subsection}{Sect.}{Sects.}
\crefname{figure}{Fig.}{Figs.}
\crefname{table}{Table}{Tables}
\crefname{equation}{Eq.}{Eq.}

\usepackage{setspace}

\newcommand{\bicol}[1]{\multicolumn{2}{c}{#1}}
\newcommand{\elec}{\electron}  
\newcommand{\Zgm}{\ensuremath{\Z/\gamma^\ast}\xspace} 

\newcommand{\taue}{$\tau_e$\xspace}
\newcommand{\taumu}{$\tau_\mu$\xspace}

\newcommand{\tauh}[1]{$\tau_{h#1}$\xspace}

\newcommand{\Zll}{\decay{\Z}{l^+l^-}}
\newcommand{\Ztautau}{\decay{\Z}{\tautau}}

\newcommand{\Zmumu}{\decay{\Z}{\mumu}}

\newcommand{\zbb}{\decay{\Z}{\bbbar}}

\newcommand{\Vj}{{\ensuremath{V\!j}}\xspace}
\newcommand{\VV}{{\ensuremath{VV}}\xspace} 

\newcommand{\HMT}{\texorpdfstring{\decay{H}{\mupm\taump}}{H -> mu tau\xspace}}
\newcommand{\mutau}{\ensuremath{\muon^\pm\tau^\mp}\xspace}
\newcommand{\hmumu}{\texorpdfstring{\ensuremath{\muon\tau_\mu}\xspace}{mumu\xspace}}
\newcommand{\hmue}{\texorpdfstring{\ensuremath{\muon\tau_e}\xspace}{mue\xspace}}
\newcommand{\hmuh}[1]{\texorpdfstring{\ensuremath{\muon\tau_{h{#1}}}\xspace}{muh#1\xspace}}
\newcommand{\mH}{{\ensuremath{m_H}}\xspace}
\newcommand{\CscBR}{\texorpdfstring{\ensuremath{\sigma(\decay{gg}{\HMT})}\xspace}{sigma(gg->H->mutau)\xspace}}

\newcommand{\BRtauX}{\ensuremath{\BR{(\decay{\tauon}{X})}}\xspace}
\newcommand{\BRHMT}{\texorpdfstring{\ensuremath{\BR{(\HMT)}}\xspace}{B(H->mutau)\xspace}}

\newcommand{\mass}{\ensuremath{m}\xspace}

\newcommand{\Retaphi}{\ensuremath{R_{\eta\phi}}\xspace}

\newcommand{\ihatpt}{\ensuremath{\hat{I}_{p_{\rm T}}}\xspace}

\newcommand{\mcorr}{\ensuremath{m_\text{corr}}\xspace}

\newcommand{\ipt}{\ensuremath{I_{p_{\rm T}}}\xspace}

\newcommand{\apt}{\ensuremath{A_{p_{\rm T}}}\xspace}
\newcommand{\dphi}{\ensuremath{\Delta\phi}\xspace}

\newcommand{\nsig}{\ensuremath{N_\text{sig}}\xspace}

\newcommand{\CLs}{\ensuremath{\text{CL}_\text{s}}\xspace}

\newcommand{\powhegbox}{\textsc{Powheg-Box}\xspace}

\begin{document}

\renewcommand{\thefootnote}{\fnsymbol{footnote}}
\setcounter{footnote}{1}


\begin{titlepage}
\pagenumbering{roman}

\vspace*{-1.5cm}
\centerline{\large EUROPEAN ORGANIZATION FOR NUCLEAR RESEARCH (CERN)}
\vspace*{1.5cm}
\noindent
\begin{tabular*}{\linewidth}{lc@{\extracolsep{\fill}}r@{\extracolsep{0pt}}}
\vspace*{-1.5cm}\mbox{\!\!\!\includegraphics[width=.14\textwidth]{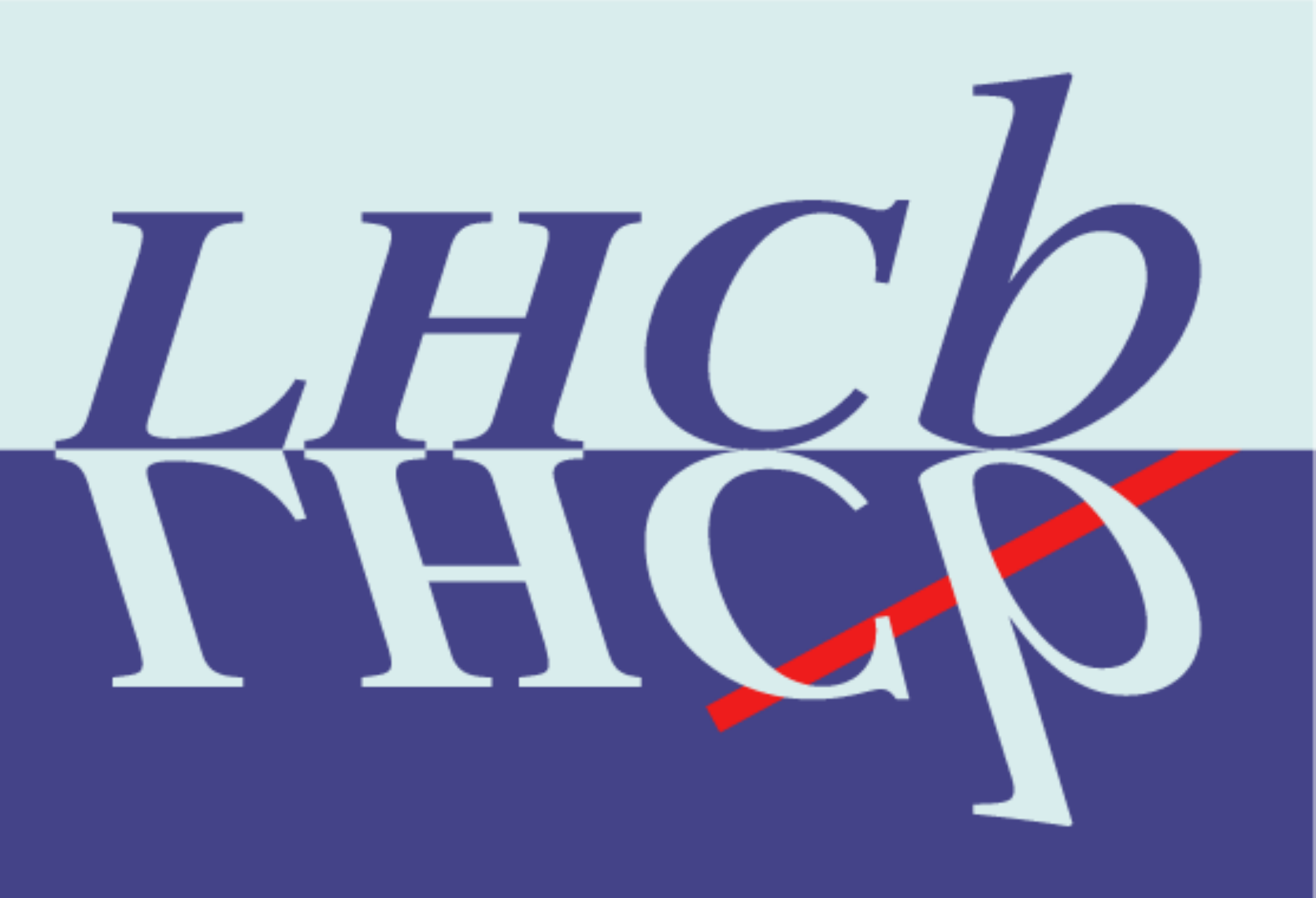}}
\\
 & & CERN-EP-2018-210 \\  
 & & LHCb-PAPER-2018-030 \\  
 & & 12 December 2018 \\ 
\end{tabular*}

\vspace*{4.0cm}

{\normalfont\bfseries\boldmath\huge
\begin{center}
  \papertitle 
\end{center}
}

\vspace*{2.0cm}

\begin{center}
\paperauthors\footnote{Authors are listed at the end of this paper.}
\end{center}

\vspace{\fill}

\begin{abstract}
  \noindent
A search is presented for a Higgs-like boson
with mass in the range 45 to 195\gevcc
decaying into a muon and a tau lepton.
The dataset consists of proton-proton interactions at a
centre-of-mass energy of 8\tev, collected by the LHCb experiment,
corresponding to an integrated luminosity of 2\invfb. 
The tau leptons are reconstructed in both leptonic and hadronic decay channels.
An upper limit on the production cross-section 
multiplied by the branching fraction at 95\% confidence level is set 
and ranges from 22\pb for a boson mass of 45\gevcc to 4\pb for a mass of 195\gevcc.

\end{abstract}

\vspace*{2.0cm}

\begin{center}
Published in Eur. Phys. J. C78 (2018) 1008
\end{center}

\vspace{\fill}

{\footnotesize 
\centerline{\copyright~\papercopyright. \href{\paperlicenceurl}{\paperlicence}.}}
\vspace*{2mm}

\end{titlepage}


\newpage
\setcounter{page}{2}
\mbox{~}
%
%
%
%

\cleardoublepage


\renewcommand{\thefootnote}{\arabic{footnote}}
\setcounter{footnote}{0}


\pagestyle{plain} 
\setcounter{page}{1}
\pagenumbering{arabic}

\section{Introduction}

Decays mediated by charged-lepton flavour-violating (CLFV) processes
are forbidden in the Standard Model (SM).
Their observation would be a clear sign for physics beyond the SM.
Such processes are predicted by several theoretical models
\cite{Blanke:2008zb,Giudice:2008uua,AguilarSaavedra:2009mx,Albrecht:2009xr,Goudelis:2011un,McKeen:2012av,Arganda:2004bz,Arganda:2014dta},
in particular those based on an effective theory with relaxed renormalisability requirements
\cite{Harnik:2012pb},
supersymmetric models
\cite{Bjorken:1977vt,DiazCruz:1999xe,Han:2000jz,Arhrib:2012ax,Arana-Catania:2013xma},
composite Higgs models \cite{Agashe:2009di,Azatov:2009na},
Randall-Sundrum models \cite{Perez:2008ee,Casagrande:2008hr},
and non-abelian flavour symmetry models \cite{Ishimori:2010au}.
Nonetheless, no evidence for CLFV effects has been reported to date.

The \lep experiments set stringent limits 
on the CLFV decay of the \Z boson~\cite{ALEPHZ, DELPHIZ, L3Z, OPALZ}.
In the presence of CLFV couplings, the decays to
\epm\mump, \epm\taump and \mump\taump
could be mediated by a Higgs boson.
At LEP2, limits on the cross-section of the 
\mbox{$\epem\to\epm\mump$}, 
\mbox{$\epem\to\epm\taump$} and 
\mbox{$\epem\to\mupm\taump$}
processes were
obtained by the \opal collaboration for centre-of-mass energies ($\sqs$)
ranging from 192 to 209\gev~\cite{Abbiendi:2001cs}.
These constraints can be translated into limits on the Higgs
CLFV decay branching fraction \cite{Harnik:2012pb,Blankenburg:2012ex},
which are on the order of $10^{-8}$
for a SM Higgs decay into an electron and muon \cite{Blankenburg:2012ex}.
Recent searches for the \HMT decay have been performed by
the \cms~\cite{Sirunyan:2017xzt} and \atlas~\cite{Aad:2015gha} collaborations
for the Higgs boson with \mbox{\mH = 125\gevcc}.
Upper limits on the branching fraction
\BRHMT have been placed by the two collaborations at 0.25\% and 1.85\%, respectively.

The possible existence of low-mass Higgs-like bosons is a feature of models
like the two-Higgs-doublet models (2HDM)~\cite{GUNION2HDM}.
Searches for such particles have been performed by the \atlas~\cite{Aad:2012cfr}
and \cms~\cite{Chatrchyan:2012vp} collaborations in the ditau decay mode.
Another scenario is that of a hidden gauge sector~\cite{Jaeckel:2010ni,Baumgart:2009tn}.
In this context, the \babar and \belle collaborations have performed searches for a resonance 
with a mass below 10\gevcc~\cite{BABARDark, BELLEDark}.
%
%
The LHCb collaboration has recently published the results of a search
for dark photons decaying into the dimuon channel, 
placing a stringent limit for the production of a dimuon
in the mass range from 10.6 to 70\gevcc~\cite{LHCb-PAPER-2017-038}.

The LHCb detector probes the forward rapidity region
which is only partially covered by the other LHC experiments, 
and triggers on particles with low transverse momenta (\pt), 
allowing the experiment to explore relatively small boson masses.
In this paper a search for CLFV decays into a muon and a tau lepton
of a Higgs-like boson with a mass ranging from 45 to 195\gevcc is presented,
using proton-proton collision data collected at \mbox{$\sqs=8$\tev}.
The Higgs-like boson is assumed to be produced by gluon-fusion,
similarly to the main production mechanism of the SM 
Higgs boson at LHC~\cite{Georgi:1977gs}.\footnote{The remaining Higgs production modes (\eg, $\sim10\%$ from Vector-Boson Fusion) are neglected in this study.}
The analysis is separated into four channels depending on the final state
of the \tauon lepton decay:
(i) single muon \decay{\taum}{\mun\neumb\neut},
(ii) single electron \decay{\taum}{\en\neueb\neut},
(iii) single charged hadron \decay{\taum}{\pim(\piz)\neut},
and (iv) three charged hadrons \decay{\taum}{\pim\pim\pip(\piz)\neut}.
They are denoted as \taumu, \taue, \tauh1, and \tauh3 respectively.
%
The main sources of background are \Ztautau decays,\footnote{
Throughout this note, \Z implies \Zgm, \ie includes contributions from
\Z boson production, virtual photon production, and also their interference.}
heavy flavour production from QCD processes (``QCD'' in the following)
and electroweak boson production accompanied by jets (``\Vj'').
%
%
This analysis utilizes reconstruction techniques and results obtained from the \Ztautau measurement by the \lhcb collaboration~\cite{LHCb-PAPER-2018-016}.


\section{Detector and simulation description}

The \lhcb detector~\cite{Alves:2008zz,LHCb-DP-2014-002} is a single-arm forward
spectrometer covering the $2< \eta <5$ pseudorapidity range,
designed for the study of particles containing \bquark or \cquark quarks. 
The detector includes a high-precision tracking system
consisting of a silicon-strip vertex detector surrounding the $pp$
interaction region, a large-area silicon-strip detector located
upstream of a dipole magnet with a bending power of $4{\mathrm{\,Tm}}$, 
and three stations of silicon-strip detectors and straw
drift tubes placed downstream of the magnet.
The tracking system provides a measurement of the momentum of charged particles with
a relative uncertainty that varies from 0.5\% at low momentum to 1.0\% at 200\gevc.
The minimum distance of a track to a primary vertex (PV), the impact parameter (IP), 
is measured with a resolution of $(15+29/\pt)\mum$,
where \pt is the component of the momentum transverse to the beam, in\,\gevc.
Photons, electrons and hadrons are identified by a calorimeter system consisting of
scintillating-pad (SPD) and preshower detectors (\presh), an electromagnetic
calorimeter (\ecal) and a hadronic calorimeter (\hcal). 
Muons are identified by a system composed of five stations of alternating layers 
of iron and multiwire proportional chambers.
%



Simulated data samples are used to calculate 
the efficiency for selecting signal processes, 
to estimate the residual background level,
and to produce templates for the fit used to determine the signal yield.
For this analysis,
the simulation is validated primarily by comparing \Zll decays in simulation and data.
The Higgs boson is generated assuming a gluon-fusion process, and with mass values
from 45 to 195\gevcc in steps of 10\gevcc,
using \mbox{\pythia8}~\cite{Sjostrand:2007gs,*Sjostrand:2006za}
with a specific \lhcb configuration~\cite{LHCb-PROC-2010-056}.
The parton density functions (PDF) are taken from the CTEQ6L set~\cite{cteq6l}.
Decays of hadronic particles are described by \evtgen~\cite{Lange:2001uf}, 
in which final-state radiation is generated using \photos~\cite{Golonka:2005pn}. 
The interaction of the particles with the detector and its response
are implemented using the \geant toolkit~\cite{Allison:2006ve, *Agostinelli:2002hh}
as described in Ref.~\cite{LHCb-PROC-2011-006}.
Samples of \HMT decays generated at next-to-leading order precision 
by \powhegbox \cite{inspire:659055,inspire:760769,inspire:845712,inspire:804159}
with the PDF set \texttt{MMHT2014nlo68cl}~\cite{inspire:1334137} 
are used for the signal acceptance determination.

%


\section{Signal selection}

This analysis uses data corresponding to a total integrated luminosity of
\mbox{${1976 \pm 23}\invpb$}~\cite{LHCb-PAPER-2014-047}.
The data collected uses a trigger system consisting of a hardware stage
followed by a software stage.
The hardware trigger requires a muon track identified by matching hits in the muon stations,
as well as a global event cut (GEC) requiring the hit multiplicity in the SPD to be less than 600.
The software trigger selects muons or electrons with a minimum \pt of 15\gevc.

The \HMT candidates are identified and reconstructed into the four channels:
\hmue, \hmuh1, \hmuh3 and \hmumu.
The \tauh3 candidates are reconstructed from the combination of three charged hadrons
from a secondary vertex (SV).
The \mutau candidates are required to be compatible with originating from a common PV.
The muon track and the tracks used to reconstruct the tau candidate 
must be in the geometrical region $2.0<\eta<4.5$.
%
%
Electron candidates are chosen amongst tracks
failing the muon identification criteria and  
falling into the acceptance of the \presh, \ecal, and \hcal sub-detectors.
A large energy deposit, $E$, in the \presh, \ecal, but not in \hcal is required,
satisfying: $E_\text{\presh} > 50\mev$,
$E_\text{\ecal}/\ptot > 0.1$, and $E_\text{\hcal}/\ptot < 0.05$, 
where \ptot is the reconstructed momentum of the electron candidate, 
after recovering the energy of the bremsstrahlung photons~\cite{brem}.
Charged hadrons are required to be in the \hcal acceptance, 
to deposit an energy $E_\text{\hcal}$ with $E_\text{\hcal}/\ptot > 0.05$, 
and to fail the muon identification criteria.
The pion mass is assigned to all charged hadrons.

%

The selection criteria need to be optimised over the \mH range used in this analysis, from 45 to 195\gevcc.
Three different sets of selection criteria are considered, 
dubbed L-selection, C-selection, and H-selection.
The C-selection is similar to that used for 
the analysis of \Ztautau decays~\cite{LHCb-PAPER-2018-016}; 
as such, it is optimised for $\mH \sim m_Z$.
The L-selection and H-selection are optimised for the \mH regions 
below and above the \Z mass respectively.
All selection sets are applied in parallel to compute background estimation and exclusion limits. 
Subsequently, for each \mH hypothesis, the chosen selection is that of L-, C-, or H-selection which provides the smallest expected signal limit,
allowing precise separation between adjacent mass regions.
%
As expected, it is found that the C-selection is optimal for a boson mass of 75 and 85\gevcc. Below and above that range the best upper limits are obtained from the L- and H-selections, respectively.
In the following discussion the requirements are applied identically
for all decay channels and selection sets unless stated otherwise.


The tau candidates are selected with $\pt>5\gevc$ for \taue,\taumu,
and $\pt>10\gevc$ for \tauh1.
For the \tauh3 candidate, the charged hadrons are required to have $\pt > 1\gevc$ and one of them
with $\pt > 6\gevc$. They are combined to form the tau candidates, which are required
to have $\pt > 12\gevc$ and an invariant mass in the range 0.7 to 1.5\gevcc.
In the H-selection, the tau candidates must have \pt in excess of 20\gevc.
This requirement is not applied in the \hmumu channel as it favours the selection of \Zmumu background.
The muon from \HMT decay is expected to have a relatively large \pt,
thus the selection requires the muon \pt to be greater than 
20\gevc, 30\gevc, and 40\gevc in the L-, C-, and H-selections, respectively.
A tighter requirement of 50\gevc is applied for the muon in the \hmumu channel in the H-selection
due to the \Zmumu background.
Additionally, for the \hmue channel, the contribution from \decay{W/Z}{\elec+\text{jet}} 
background is suppressed by requiring the transverse momentum of the muon to be larger
than that of the \taue candidate.


The relatively large lifetime of the \tauon lepton is used to suppress prompt background.
For the \tauh3 candidate, a SV is reconstructed.
A correction to the visible invariant mass, \mass, 
computed from the three-track combination, 
is obtained by exploiting the direction of flight defined from the PV to the SV.
The relation used is \mbox{$\mcorr = \sqrt{m^2 + p^2 \sin^2\theta} + p\sin\theta$},
where $\theta$ is the angle between the momentum of the \tauh3 candidate,
and its flight direction.
The \mcorr value is required to not exceed 3\gevcc.
A time-of-flight variable is also computed from the distance of flight
and the partially reconstructed momentum of the \tauon lepton,
and a minimum value of 30\fs is required.
The \mcorr and time-of-flight requirements together retain 80\% of the signal,
while rejecting about 75\% of the QCD background.
For tau decay channels with a single charged particle, 
it is not possible to reconstruct a SV, 
and a selection on the particle IP is applied.
A threshold of IP $>10\mum$ selects 85\% of the \taue and \tauh1 candidates,
and rejects about 50\% of the \Vj background.
The threshold is increased to $50\mum$ for \taumu candidates,
in order to suppress \Zmumu background.
The prompt muon instead is selected by requiring IP less than 50\mum,
allowing up to 50\% rejection of QCD and \Ztautau backgrounds.


The two leptons from the Higgs decay should be approximately back-to-back 
in the plane transverse to the beam.
The absolute difference in azimuthal angle of muon and tau candidates 
is required to be greater than 2.7 radians.
This rejects 50\% of the \Vj background.
The transverse momentum asymmetry of the two particles, defined as
\mbox{$\apt = {|\pt{}_1 - \pt{}_2|}/{(\pt{}_1 + \pt{}_2})$},
can be used to effectively suppress various background processes.
The background from the \Vj processes is suppressed by up to 60\% for
the \hmuh1 channel by requiring $\apt < 0.4\;(0.5)$ in the L-selection (S-selection), 
because of the large \pt imbalance between the high-\pt muon from the vector boson
and a hadron from a jet.
For the \hmue channel,
the worse momentum resolution increases the average \apt value, 
hence a softer selection $\apt < 0.6$ is used to preserve efficiency.
On the contrary, for the \hmumu channel, a tighter cut is applied 
to suppress the dominant background from \Zmumu decays. 
By requiring $\apt > 0.3\ (0.4)$ in the L-selection and C-selection (H-selection),
such background is reduced by 80\%, while the signal decreases to 70\%.


The two leptons from the Higgs decay are required to be 
isolated from other charged particles.
Two particle-isolation variables are defined as 
\mbox{$\ipt = (\vec{p}_\text{cone})_\text{T}$}
and
\mbox{$\ihatpt = {\pt}/{(\vec{p} + \vec{p}_\text{cone})}_\text{T}$}
where $\vec{p}$ is the momentum of the lepton candidate,
the subscript $\text{T}$ denotes the component in the transverse plane,
and $\vec{p}_\text{cone}$ is the sum of the momenta of all charged tracks
within a distance \mbox{\Retaphi = 0.5} in the $(\eta,\phi)$ plane around the lepton candidate.
The isolation requirement $\ihatpt>0.9$ is applied 
to the muon and tau candidates for all decay channels and selection sets,
and retain 70\% of the signal candidates while rejecting 90\% of QCD events.
In addition, a cut $\ipt < 2\gevc$ is applied in the L-selection to both candidates,
as the lower \pt reduces the background rejection power of the \ihatpt variable.


The selection criteria common or specific to each selection set and decay channel 
are summarised in \cref{tab:sel}.
The signal selection efficiencies are found to vary from 10 to 50\%.
Due to the kinematic selection, the decay channels are mutually exclusive
and just one \mupm\taump candidate per event is found.

\begin{table}
  \caption{Requirements for each decay channel and selection set.}
  \label{tab:sel}
\centering
\small
\begin{tabular}{llllll}
\toprule
Selection set    & Variable             & \hmue  & \hmuh1 & \hmuh3 & \hmumu \\
\midrule
All &\pt{(\tauon)}          [\!\gevc{}] & $>5$      & $>10$     & $>12$     & $>5$ \\
  &$\pt{(\tau_{h3}^\text{prong1})}$ [\!\gevc{}] &--- & ---       & $>1$& --- \\
  &$\pt{(\tau_{h3}^\text{prong2})}$ [\!\gevc{}] &--- & ---       & $>1$& --- \\
  &$\pt{(\tau_{h3}^\text{prong3})}$ [\!\gevc{}] &--- & ---       & $>6$& --- \\
  &$\pt{(\mu)}-\pt{(\tau)}$ [\!\gevc{}] & $>0$      & ---       & ---       & --- \\
  &\mass{(\tauh3)}         [\!\gevcc{}] & ---       & ---       & 0.7--1.5  & --- \\
  &\mcorr{(\tauh3)}        [\!\gevcc{}] & ---       & ---       & $>3$      & --- \\
  &Time-of-flight (\tauh3)    [\!\fs{}] & ---       & ---       & $>30$     & --- \\
  &IP(\tauon)                [\!\mum{}] & $>10$     & $>10$     & ---       & $>50$ \\
  &IP(\muon)                 [\!\mum{}] & $<50$     & $<50$     & $<50$     & $<50$ \\ 
  &\dphi                     [\!\rad{}] & $>2.7$    & $>2.7$    & $>2.7$    & $>2.7$ \\
  &\ihatpt{(\tauon)}                    & $>0.9$    & $>0.9$    & $>0.9$    & $>0.9$ \\
  &\ihatpt{(\muon)}                     & $>0.9$    & $>0.9$    & $>0.9$    & $>0.9$ \\
\midrule
L-selection & \pt{(\muon)}   [\!\gevc{}] & $>20$  & $>20$  & $>20$ & $>20$  \\
            & \apt                       & $<0.6$ & $<0.4$ & ---   & $>0.3$ \\
            & \ipt{(\tauon)} [\!\gevc{}] & $<2$   & $<2$   & $<2$  & $<2$   \\
            & \ipt{(\muon)}  [\!\gevc{}] & $<2$   & $<2$   & $<2$  & $<2$   \\
\midrule
C-selection & \pt{(\muon)}  [\!\gevc{}] & $>30$ & $>30$  & $>30$ & $>30$  \\
            & \apt                      & ---   & $<0.5$ & ---   & $>0.3$ \\
\midrule
H-selection & \pt{(\tauon)} [\!\gevc{}] & $>20$ & $>20$ & $>20$ & ---    \\
            & \pt{(\muon)}  [\!\gevc{}] & $>40$ & $>40$ & $>40$ & $>50$  \\
            & \apt                      & ---   & ---   & ---   & $>0.4$ \\
\bottomrule
\end{tabular}
\end{table}


\section{Background estimation}\label{sec:background}

\begin{figure}[!t]
\centering
\mbox{
  \centering
  \subfigure{\includegraphics[height=37mm]{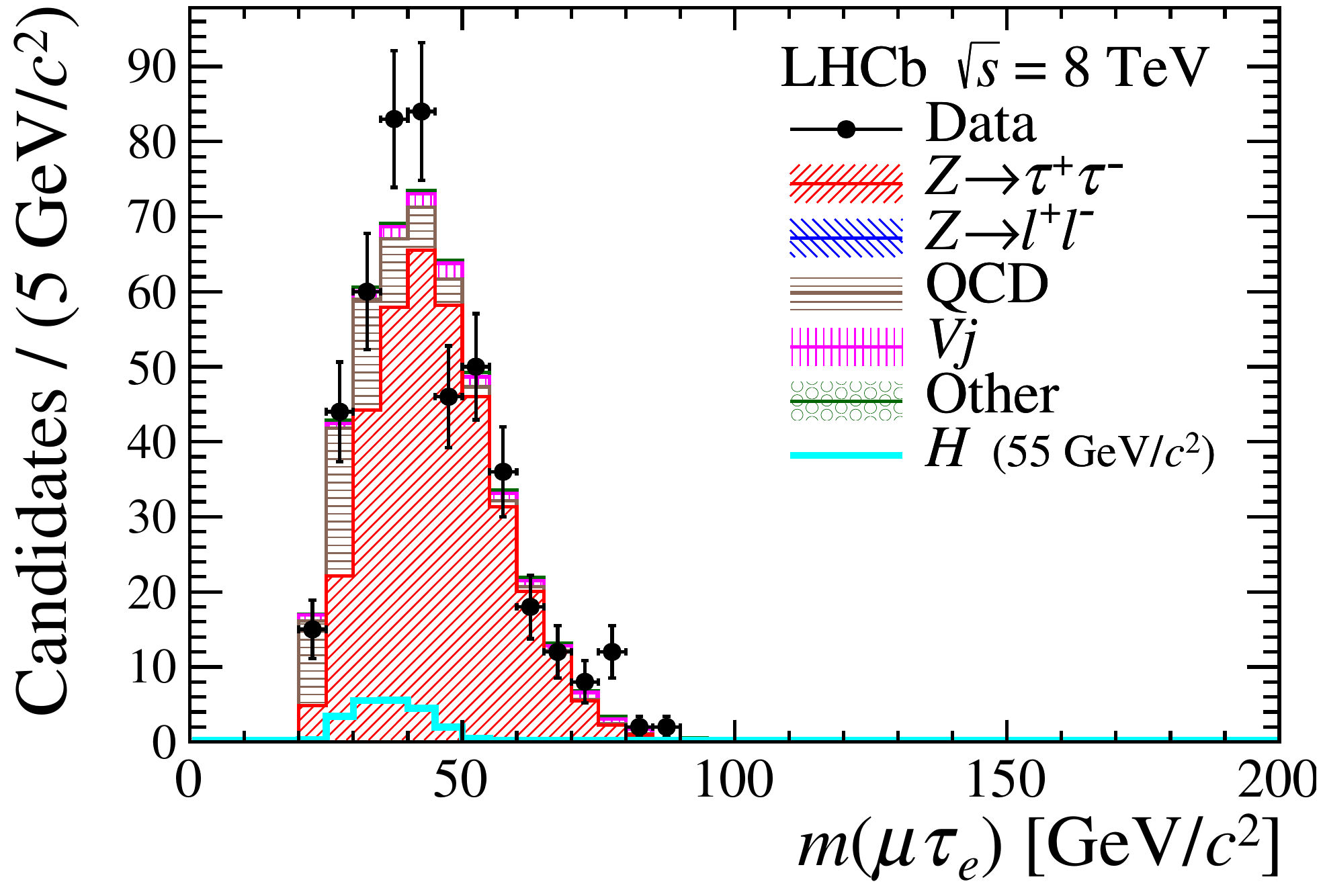}}
  \subfigure{\includegraphics[height=37mm,trim={16mm 0 0 0},clip]{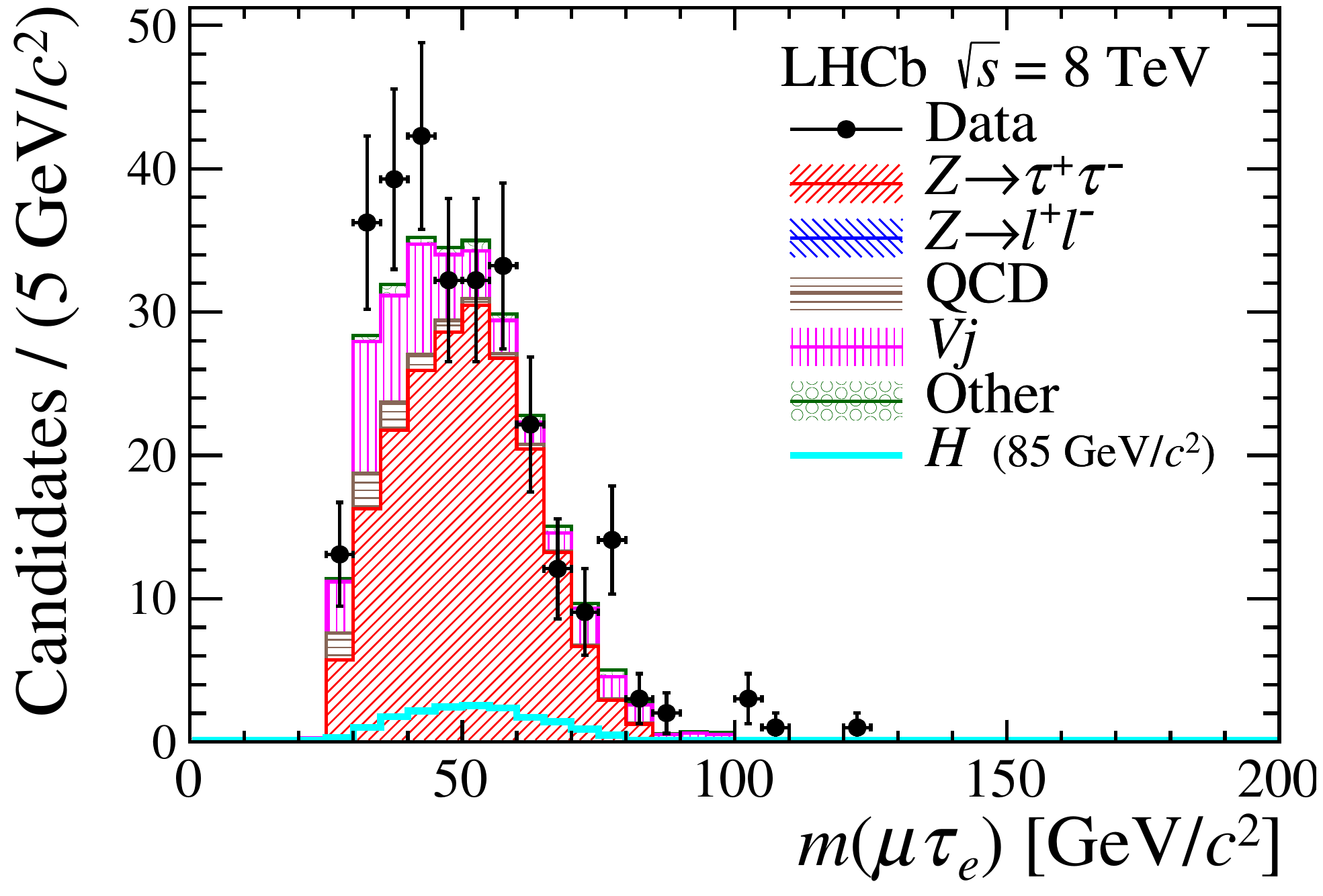}}
  \subfigure{\includegraphics[height=37mm,trim={16mm 0 0 0},clip]{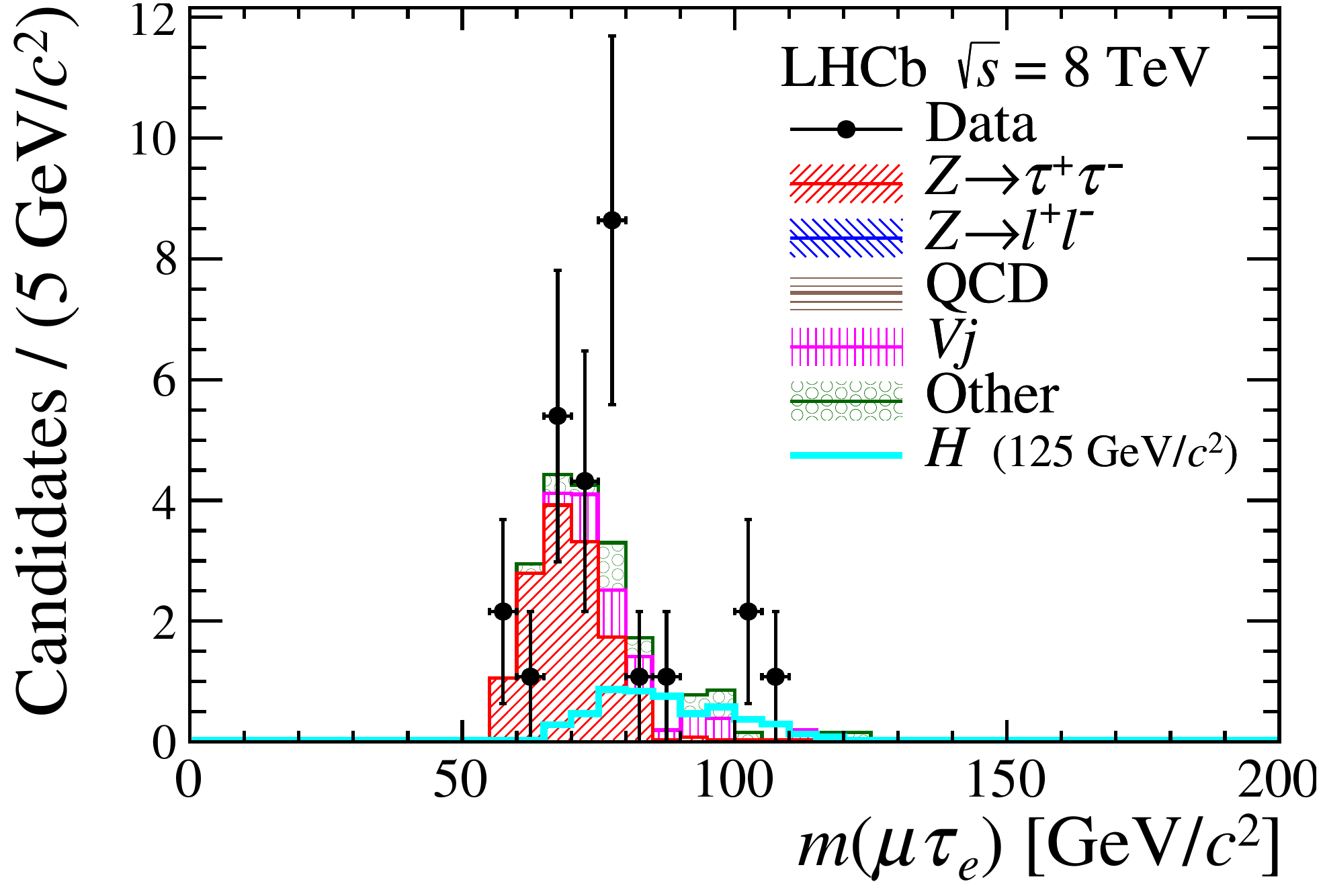}}
}\vspace*{-10pt}
\mbox{
  \centering
  \subfigure{\includegraphics[height=37mm]{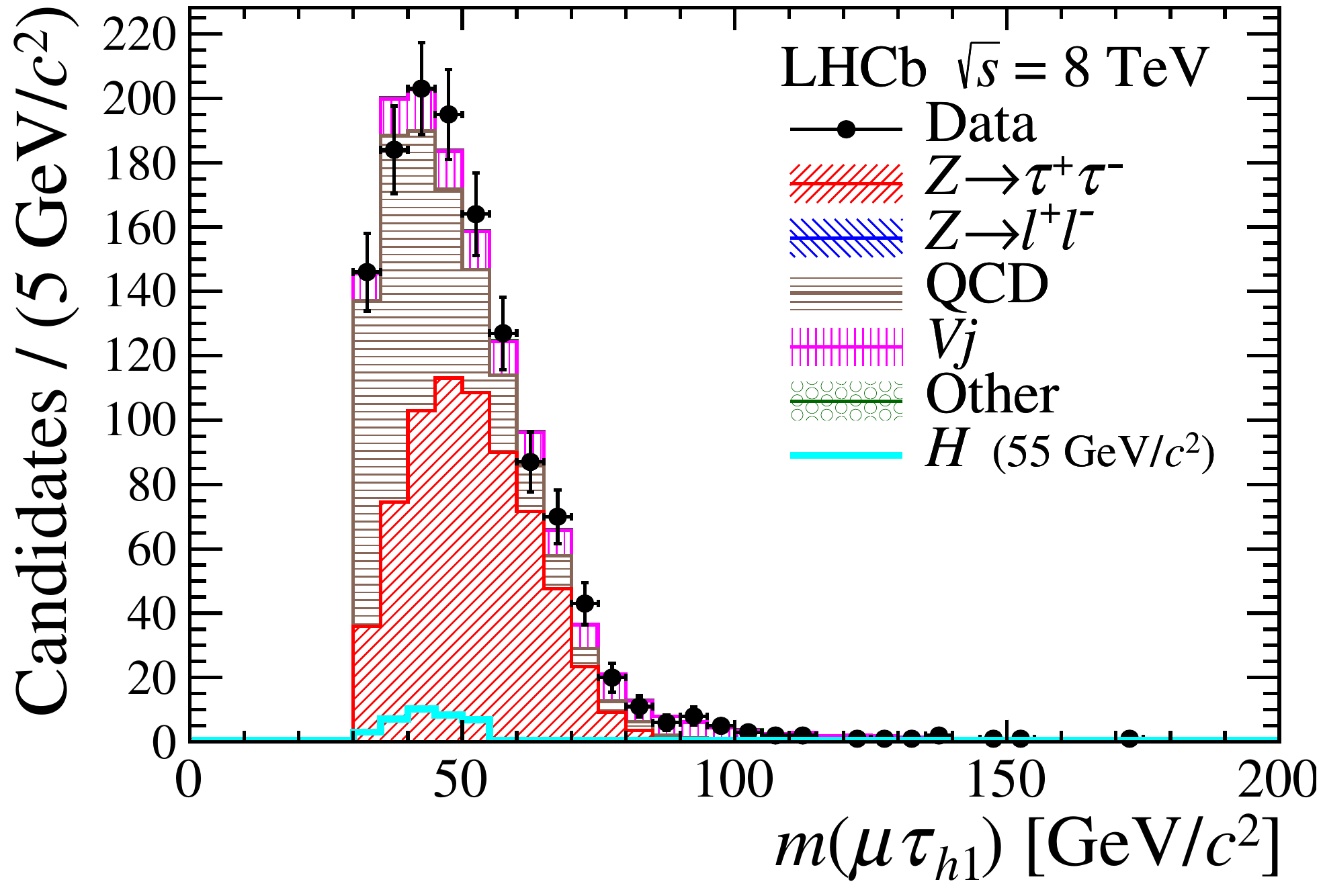}}
  \subfigure{\includegraphics[height=37mm,trim={16mm 0 0 0},clip]{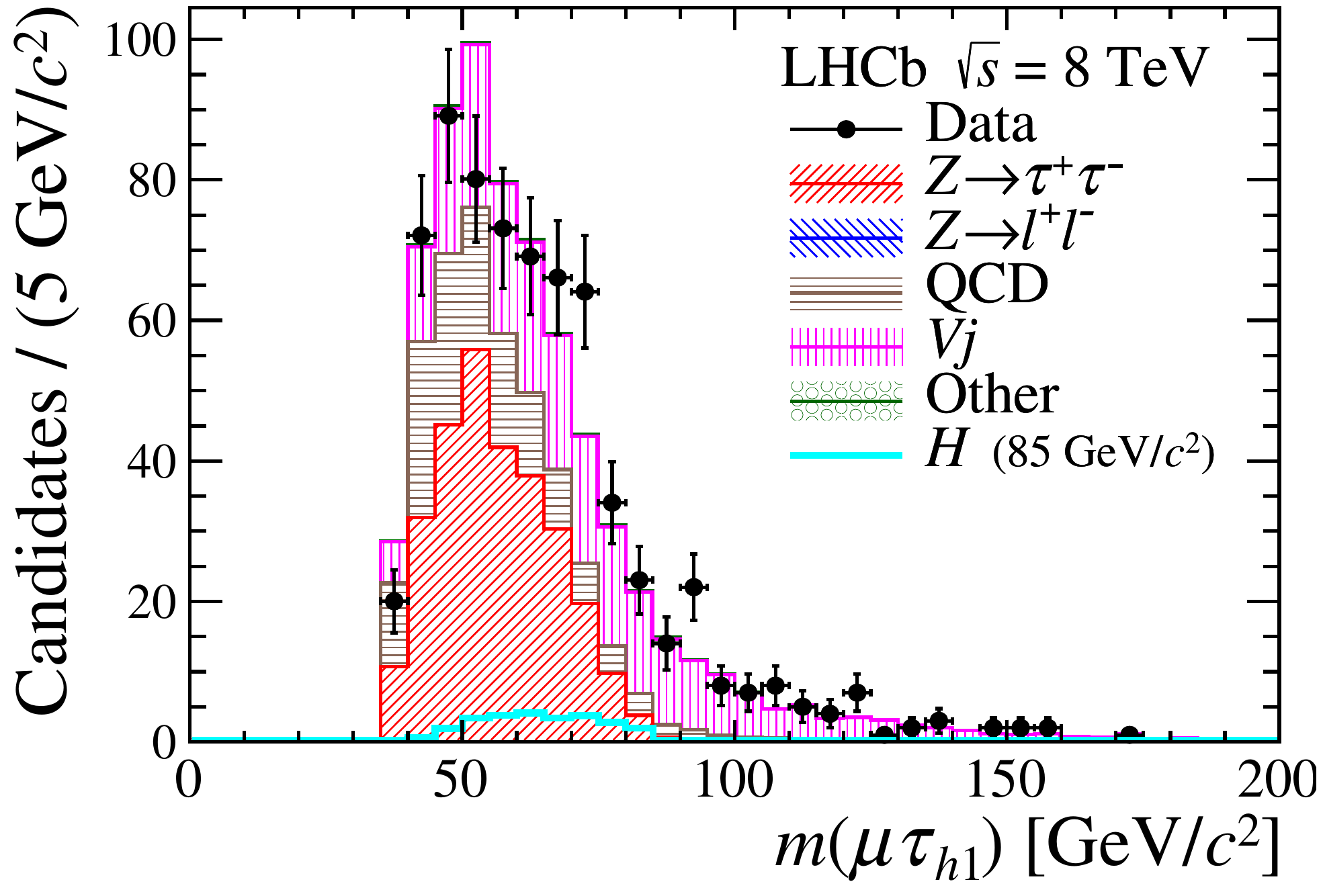}}
  \subfigure{\includegraphics[height=37mm,trim={16mm 0 0 0},clip]{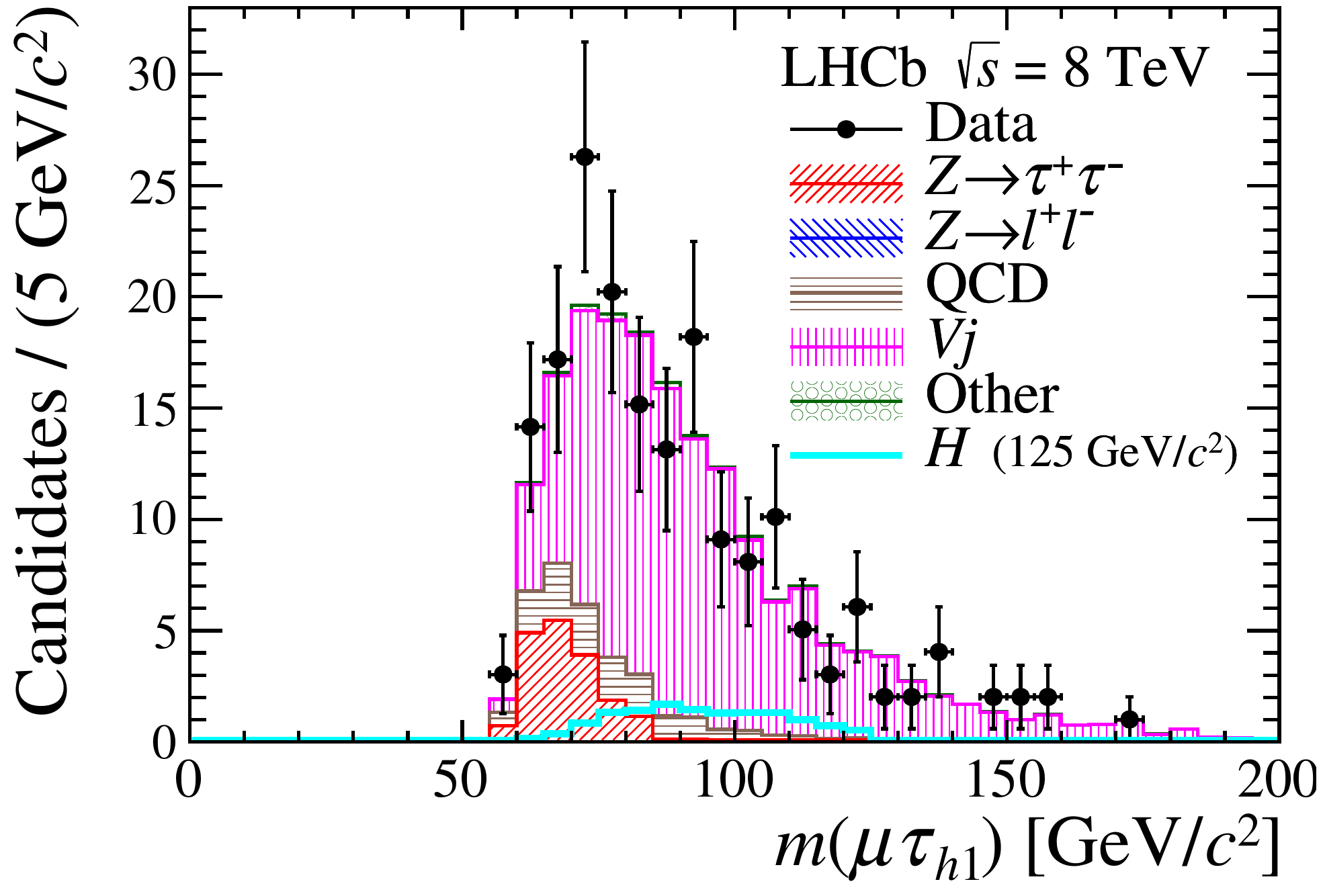}}
}\vspace*{-10pt}
\mbox{
  \centering
  \subfigure{\includegraphics[height=37mm]{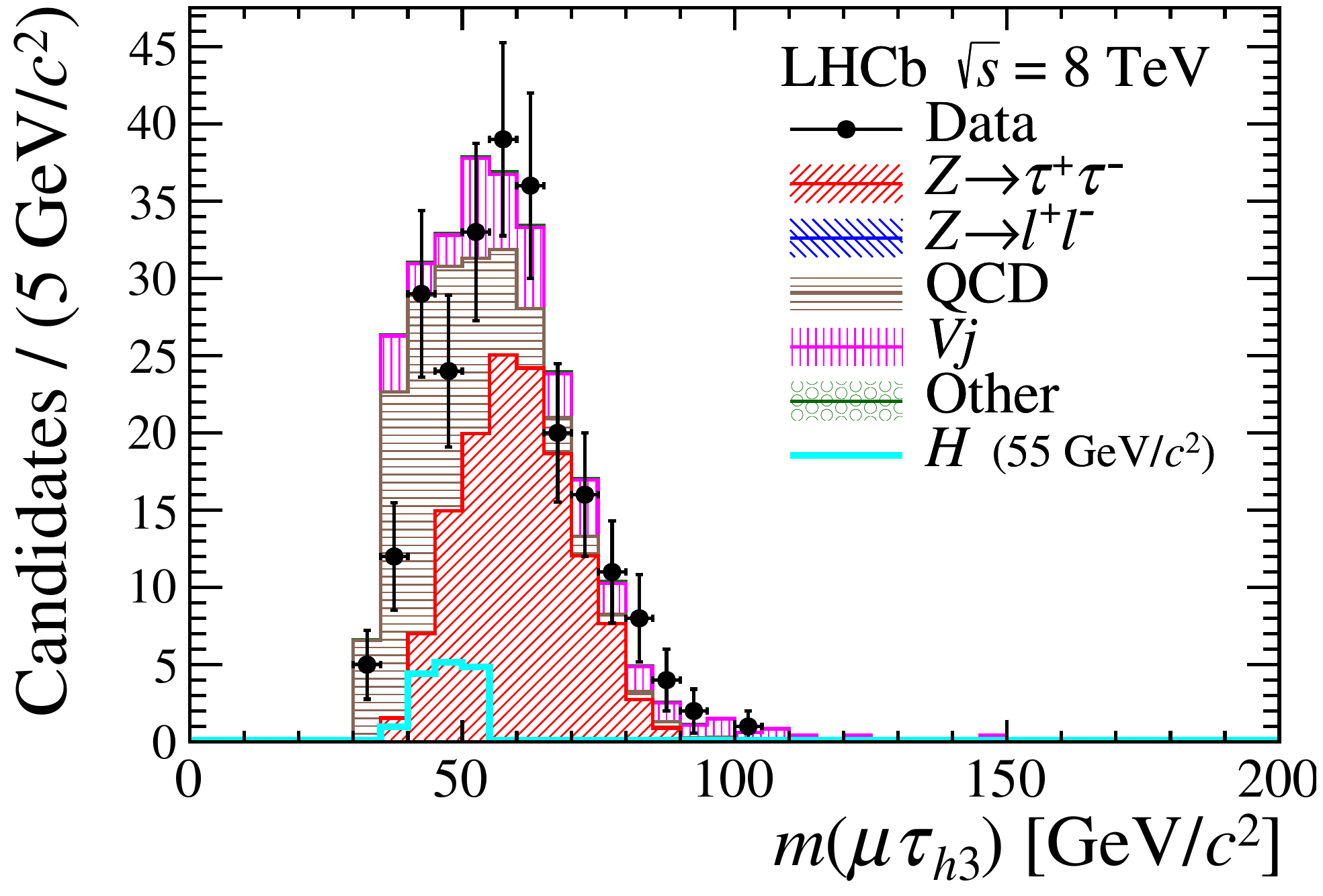}}
  \subfigure{\includegraphics[height=37mm,trim={16mm 0 0 0},clip]{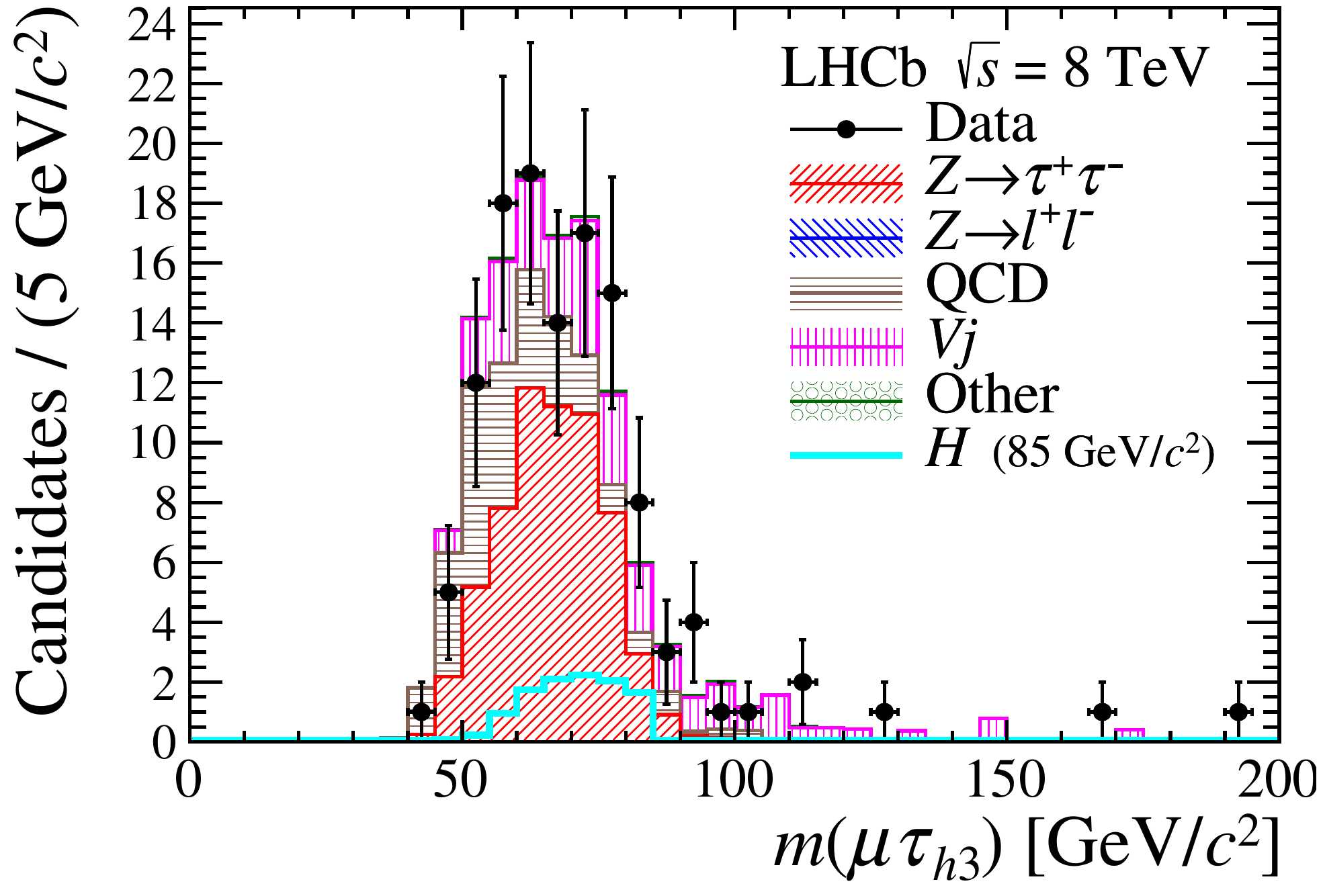}}
  \subfigure{\includegraphics[height=37mm,trim={16mm 0 0 0},clip]{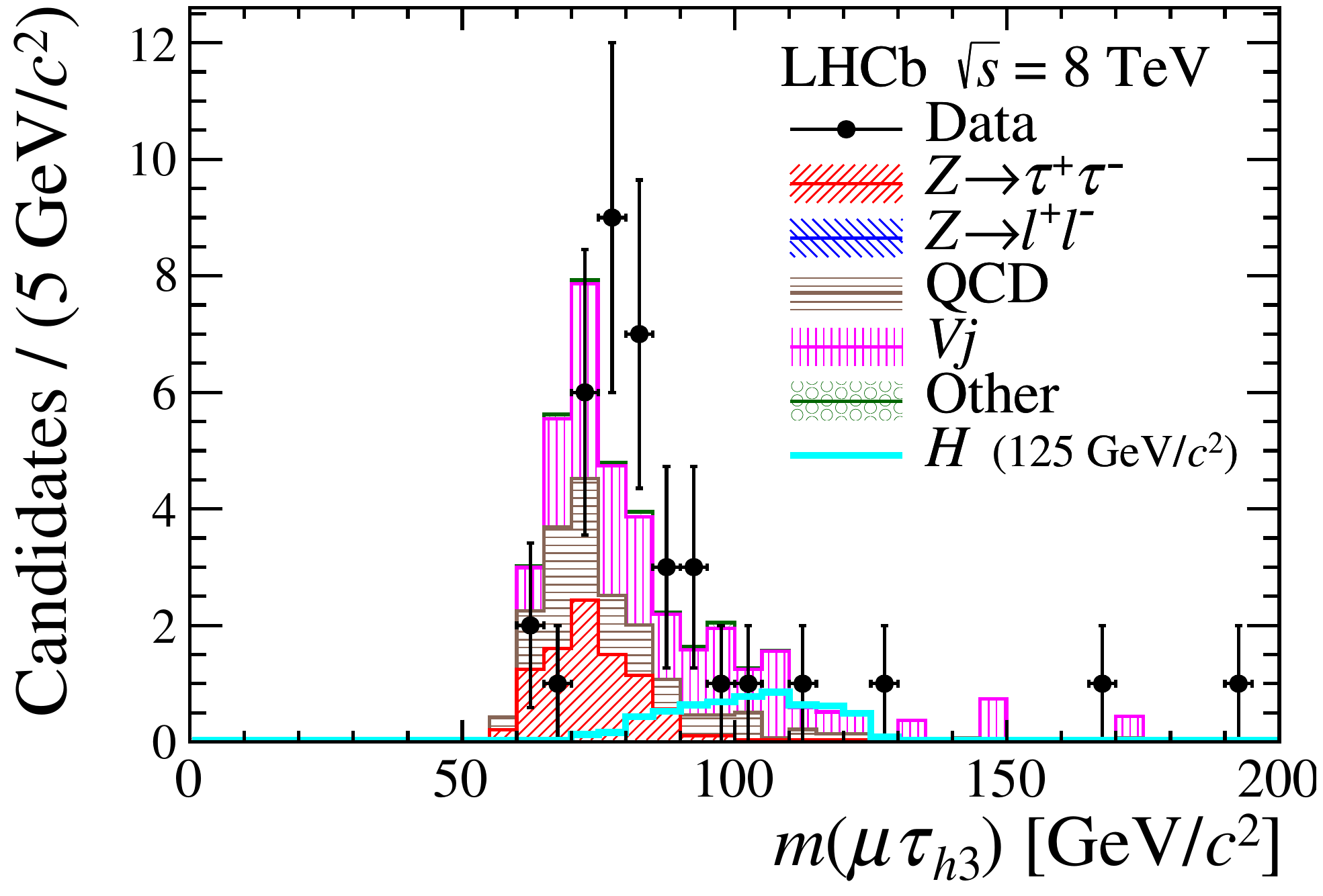}}
}\vspace*{-10pt}
\mbox{
  \centering
  \subfigure{\includegraphics[height=37mm]{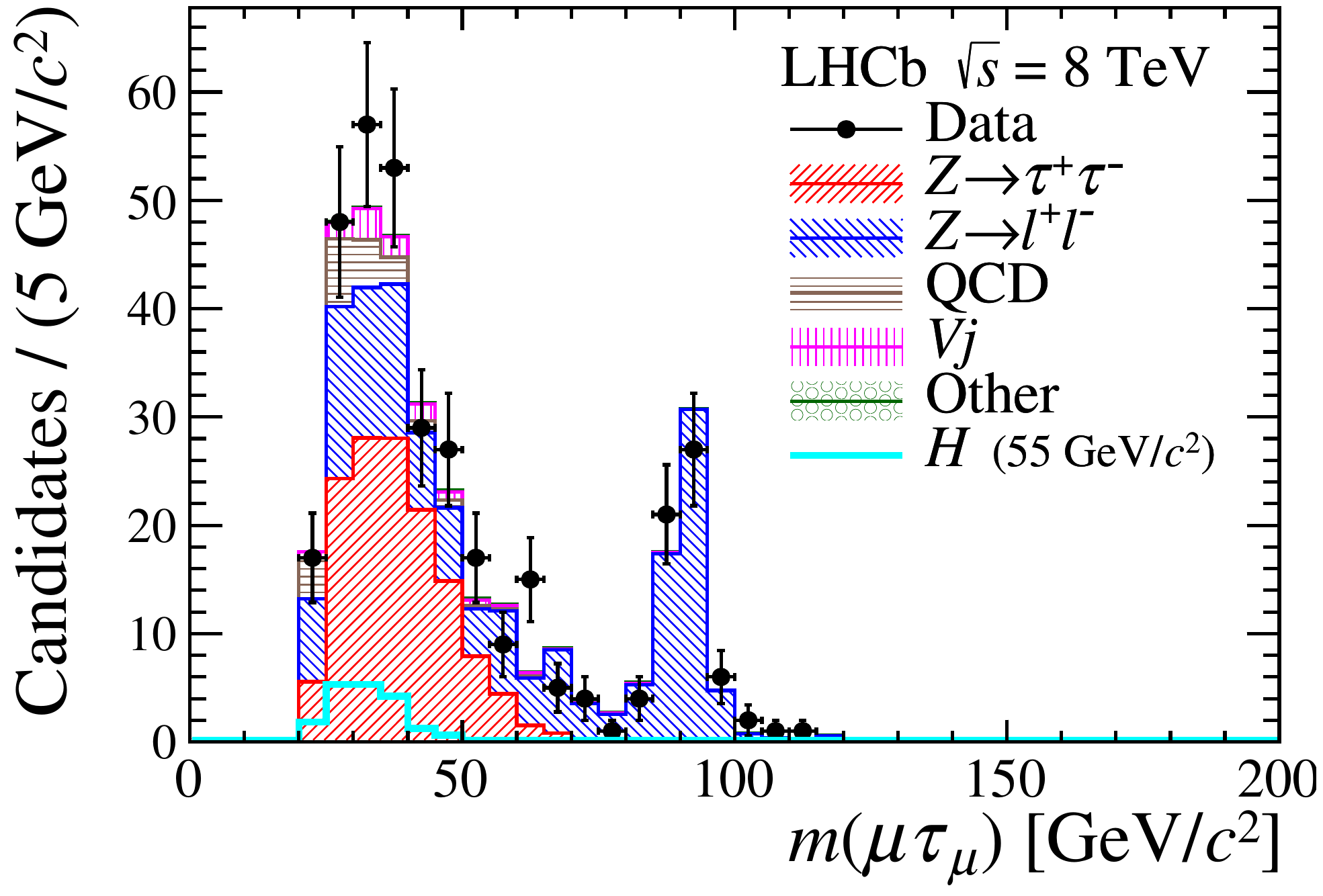}}
  \subfigure{\includegraphics[height=37mm,trim={16mm 0 0 0},clip]{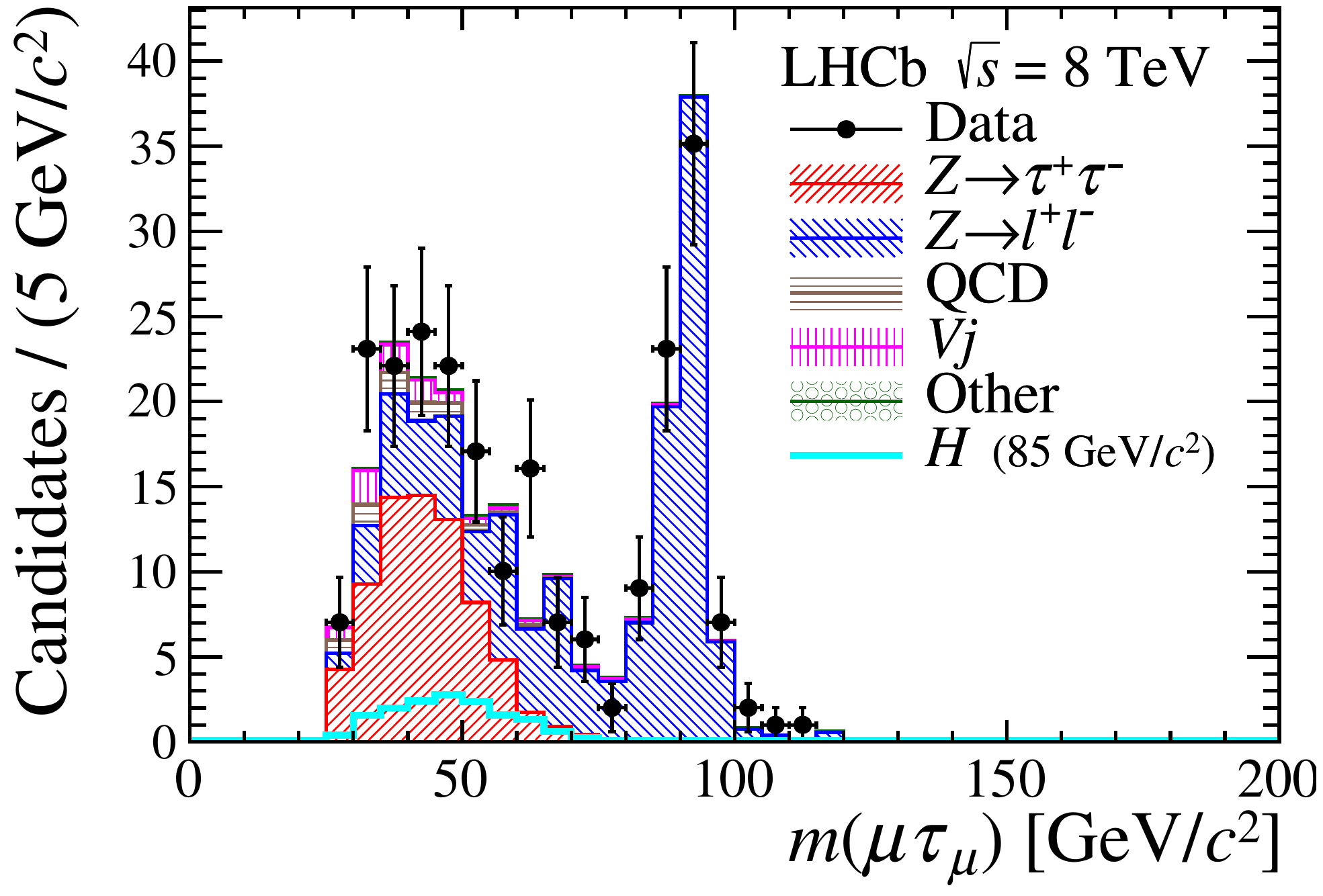}}
  \subfigure{\includegraphics[height=37mm,trim={16mm 0 0 0},clip]{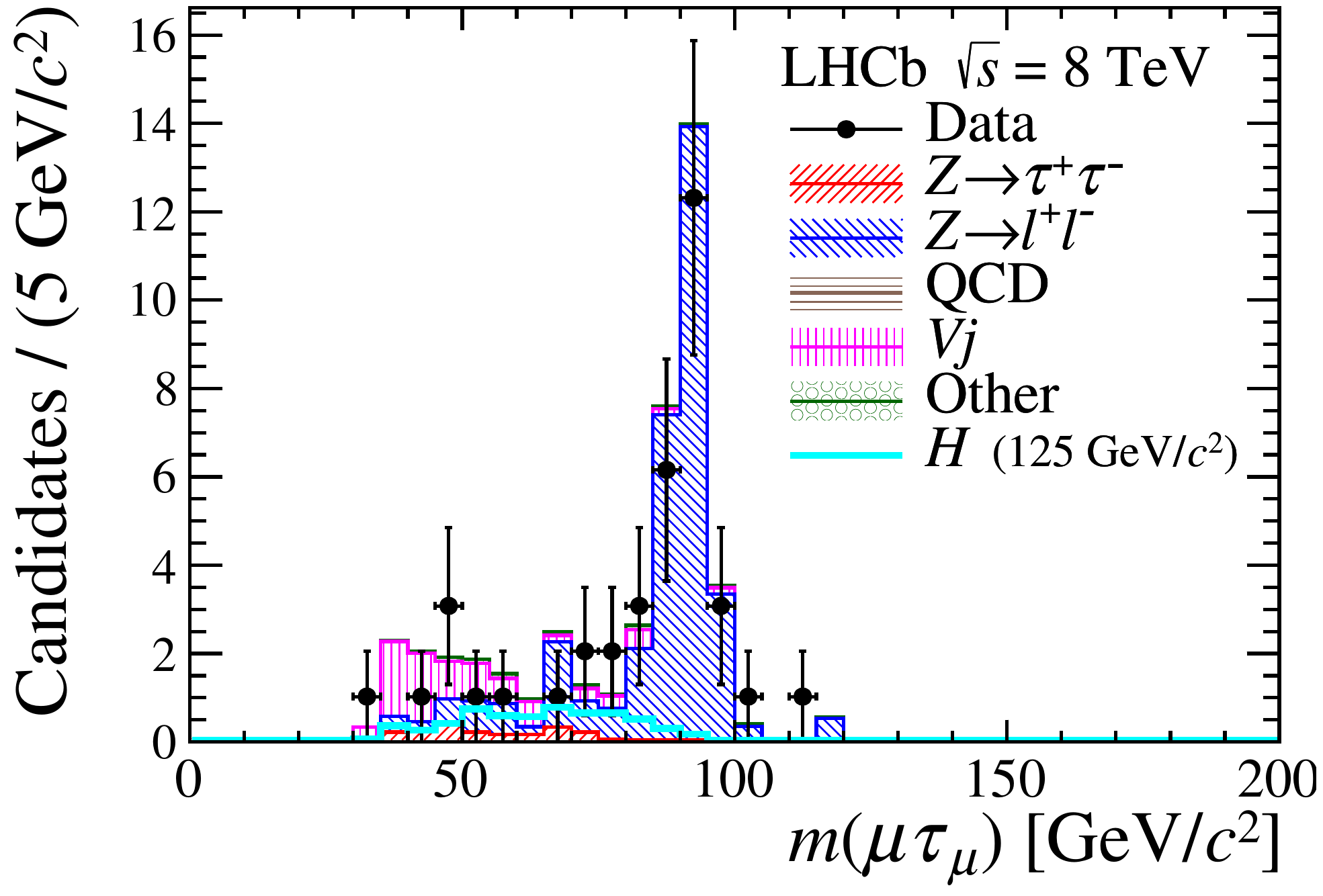}}
}
\caption{\label{fig:backgrounds}
Invariant-mass distributions for the \mutau candidates for the four decay channels
(from top to bottom: \hmue, \hmuh1, \hmuh3, \hmumu) and the three selections
(from left to right: L-selection, C-selection, H-selection).
The distribution of candidates observed (black points) is compared with 
backgrounds (filled colour, stacked), and with signal hypothesis (cyan).
The signal is normalised to $\sqrt{N}$,
with $N$ the total number of candidates in the corresponding data histogram.
}
\end{figure}

Several background processes are considered: \Ztautau, \Zll ($l=\elec,\muon)$, QCD, 
\Vj, double bosons production (\VV), \ttbar, and \zbb.
All backgrounds except \Ztautau are estimated 
following the procedures described in Ref.~\cite{LHCb-PAPER-2018-016}.
The expected yields can be found in \cref{tab:backgrounds}.
The corresponding invariant-mass distributions
compared with candidates observed in the data are shown in \cref{fig:backgrounds}.
For illustration, examples of \HMT distributions from simulation are also superimposed.

The \Ztautau background is estimated from the cross-section 
measured by the LHCb collaboration~\cite{LHCb-PAPER-2018-016} 
where the reconstruction efficiency is determined from data,
and the acceptance and selection efficiency are obtained from simulation.
The estimated background includes a small amount of cross-feed from different 
final states of the tau decay, as determined from simulation.
The \Zmumu background is dominant in the \hmumu channel.
The corresponding invariant-mass distribution is obtained from simulation 
and normalised to data in the \Z peak region, from 80 to 100\gevcc.
In order to suppress the potential presence of signal in this region, 
the muons are required to be promptly produced.
For other channels, the \Zll decay becomes a background source in case a lepton is misidentified.
This contribution is computed from the \Zll in data, 
and weighted by the particle misidentification probability obtained from simulation.

The QCD and \Vj backgrounds are inferred from data using the same 
criteria as for the signal but selecting same-sign \mupm\taupm candidates.
Their amounts are determined by a fit to the 
distribution of $\pt{(\muon)}-\pt{(\tauon)}$, 
with templates representing each of them. 
The template for the QCD component is obtained from data 
requiring an anti-isolation $\ihatpt < 0.6$ selection. 
The distribution obtained from simulation is used for the \Vj component.
Factors are subsequently applied for the correction of the relative yield 
of opposite-sign to same-sign candidates. 
For the QCD background the number of anti-isolated opposite-sign candidates 
found in data is used in the calculation of the correction factor,
where it is found to be close to unity.
The factors are found consistent with the simulation.
The factors for the \Vj component are taken from simulation, and are in general larger than unity
(1.3 for \hmue up to 3.1 for \hmuh1, for the L-selection).
The minor contributions from \VV, \ttbar, and \zbb processes are estimated from simulation.

\begin{table}
\caption{\label{tab:backgrounds}
Expected number of background candidates from each component, 
total background with uncertainty, 
and number of observed candidates with statistical uncertainty,
from each decay channel and selection set.
}
\centering
\scalebox{0.95}{
\begin{tabular}{
l
l
r@{\small$\,\pm\,$}>{\small}r
r@{\small$\,\pm\,$}>{\small}r
r@{\small$\,\pm\,$}>{\small}r
r@{\small$\,\pm\,$}>{\small}r
}
\toprule
Selection set& Process           & \bicol{\hmue} & \bicol{\hmuh1} & \bicol{\hmuh3} & \bicol{\hmumu}\\
\midrule
L-selection & \Ztautau        &  371.1 & 26.0 &   681.7 & 47.1 &  135.1 & 11.7 &  137.4 &  9.5 \\
        & \Zll             &    8.2 &  1.6 &     4.0 &  1.8 &   \bicol{---} &  155.3 &  5.0 \\
        & QCD              &   67.5 & 10.6 &   463.6 &  5.4 &   93.1 & 10.9 &   19.4 &  5.5 \\
        & \Vj              &   14.5 & 10.3 &   143.2 & 58.6 &   40.1 & 15.8 &   10.7 &  5.8 \\
        & VV               &    3.4 &  0.3 &     0.9 &  0.2 &    0.3 &  0.1 &    0.3 &  0.1 \\
        & \ttbar           &    1.7 &  0.1 &     1.3 &  0.1 &    0.7 &  0.1 &    1.3 &  0.2 \\
        & \zbb             &    0.2 &  0.2 &     0.2 &  0.2 &    0.1 &  0.1 &    0.2 &  0.2 \\
\cmidrule{2-10}
        & Total background &  466.6 & 28.0 &  1294.9 & 75.5 &  269.4 & 20.3 &  324.5 & 12.5 \\
\cmidrule{2-10}
        & Observed         &  472.0 & 21.7 &  1284.0 & 35.8 &  240.0 & 15.5 &  344.0 & 18.5 \\
\midrule
C-selection & \Ztautau         &  200.0 & 14.3 &   288.1 & 20.2 &   61.3 &  5.5 &   71.7 &  5.2 \\
        & \Zll             &    8.0 &  1.7 &     4.3 &  1.8 &   \bicol{---} &  126.7 &  4.5 \\
        & QCD              &   10.0 & 14.0 &   137.9 & 14.0 &   29.9 &  9.0 &    6.1 &  3.6 \\
        & \Vj              &   48.3 & 17.2 &   242.9 & 25.3 &   30.8 & 17.6 &    7.9 &  4.7 \\
        & VV               &    3.4 &  0.3 &     1.5 &  0.2 &    0.3 &  0.1 &    0.3 &  0.1 \\
        & \ttbar           &    2.5 &  0.1 &     1.6 &  0.1 &    0.7 &  0.1 &    1.5 &  0.2 \\
        & \zbb             &    0.1 &  0.1 &     0.1 &  0.1 &    0.1 &  0.1 &    0.1 &  0.1 \\
\cmidrule{2-10}
        & Total background &  272.3 & 17.8 &   676.4 & 35.2 &  123.1 & 15.0 &  214.3 &  8.1 \\
\cmidrule{2-10}
        & Observed         &  296.0 & 17.2 &   679.0 & 26.1 &  123.0 & 11.1 &  235.0 & 15.3 \\
\midrule
H-selection & \Ztautau       &   13.7 &  1.8 &    18.4 &  1.6 &    8.9 &  1.1 &    2.2 &  0.4 \\
          & \Zll           &    4.7 &  1.1 &     2.5 &  1.1 &   \bicol{---} &   33.7 &  2.3 \\
          & QCD            &   \bicol{---} &    15.8 &  6.3 &    9.7 &  5.1 &   \bicol{---} \\
          & \Vj            &    3.5 &  2.6 &   142.6 & 26.0 &   18.6 & 16.5 &    7.8 &  4.0 \\
          & VV             &    1.7 &  0.2 &     1.0 &  0.2 &    0.1 &  0.1 &    0.2 &  0.1 \\
          & \ttbar         &    1.2 &  0.1 &     0.9 &  0.1 &    0.4 &  0.1 &    0.8 &  0.1 \\
          & \zbb           &    0.1 &  0.1 &     0.1 &  0.1 &    0.1 &  0.1 &    0.1 &  0.1 \\
\cmidrule{2-10}
        & Total background &   24.9 &  3.4 &   181.2 & 26.7 &   37.8 & 13.6 &   44.7 &  4.6 \\
\cmidrule{2-10}
        & Observed         &   27.0 &  5.2 &   184.0 & 13.6 &   37.0 &  6.1 &   39.0 &  6.2 \\
\bottomrule
\end{tabular}
}
\end{table}


\section{Results}

The signal cross-section multiplied by the branching fraction is given by
\begin{equation}\label{eq:cscbr}
\CscBR = \nsig / (\lum\cdot\BRtauX\cdot\varepsilon), 
\end{equation}
where \nsig is the signal yield obtained from the fit procedure described below,
\lum the total integrated luminosity, 
\BRtauX the tau branching fraction,
and $\varepsilon$ the detection efficiency.
The latter is the product of acceptance,
reconstruction, and offline selection efficiencies. 
These efficiencies are obtained from simulated samples and data
for each decay channel and selection set, following the methods 
developed for the \Ztautau measurement~\cite{LHCb-PAPER-2018-016}.
The acceptance obtained from the \powhegbox generator is identical 
for the \hmue, \hmuh3, and \hmumu channels, 
varying from 1.0\% for \mbox{$\mH = 195\gevcc$}
to 3.2\% for \mbox{$\mH = 75\gevcc$}.
The reconstruction efficiency, which is the product of contributions from trigger, tracking, and particle identification, is in the range 40--70\%, 
but only about 15\% in the case of the \hmuh3 channel
because of the limited tracking efficiency for the low-momentum hadrons.
With the exception of the \hmumu channel, 
the selection efficiency is 18--30\% in the L-selection,
and 24--49\% in the C-selection and H-selection.
In the case of the \hmumu channel,
the tighter selection on the muon \pt and impact parameter 
reduces the selection efficiency to 10--15\%.

The systematic uncertainties are summarised in \cref{tab:uncer}.
The uncertainty on the acceptance receives 
contributions from the gluon PDF uncertainty,
as well as from factorization and renormalisation scales.
The uncertainties on the reconstruction and selection efficiencies
are estimated from simulation and are calibrated using data as
described in Ref.~\cite{LHCb-PAPER-2018-016}.
The uncertainty associated with the invariant-mass shape is handled by
selecting the weakest expected limits among the different choices of distribution
(kernel estimation and histograms with different bin widths are used).
The uncertainties on the integrated luminosity and acceptance are fully correlated among channels, 
while only a partial correlation is found for the reconstruction efficiency uncertainties.
All the other uncertainties are taken as uncorrelated.

\begin{table}[t]
\caption{\label{tab:uncer}
Relative systematic uncertainties (in \%) on the normalisation factors
in the cross-section calculation. 
When the uncertainty depends on \mH a range is indicated.
}
\centering
\scalebox{0.9}{
\begin{tabular}{lcccc}
\toprule
 & \hmue & \hmuh1 & \hmuh3 & \hmumu\\
\midrule
Luminosity                        & 1.16     & 1.16     & 1.16     & 1.16\\
Tau branching fraction            & 0.22     & 0.18     & 0.48     & 0.23\\
PDF                               & 2.6--7.1 & 3.5--7.2 & 2.6--7.3 & 3.0--7.9\\
Scales                            & 0.9--1.9 & 0.8--1.7 & 0.9--1.7 & 0.9--1.9\\
Reconstruction efficiency         & 1.8--3.6 & 1.9--5.4 & 3.3--7.1 & 1.5--3.3\\
Selection efficiency              & 2.5--6.0 & 1.9--4.1 & 4.0--9.3 & 3.8--8.5\\
\bottomrule
\end{tabular}
}
\end{table}

The signal yield is determined from a simultaneous extended likelihood fit
of the binned invariant-mass distributions of the \muon\tauon candidates.
The distributions for signal are obtained from simulation,
while distributions of the different background sources are obtained using the
method described in \cref{sec:background}. 
The amount of each background component as well as other terms in \cref{eq:cscbr}
containing uncertainties are treated as nuisance parameters and are constrained to 
a Gaussian distribution with mean and standard deviation corresponding to
the expected value and its uncertainty, respectively.

\begin{figure}
\centering
\includegraphics[width=0.6\textwidth]{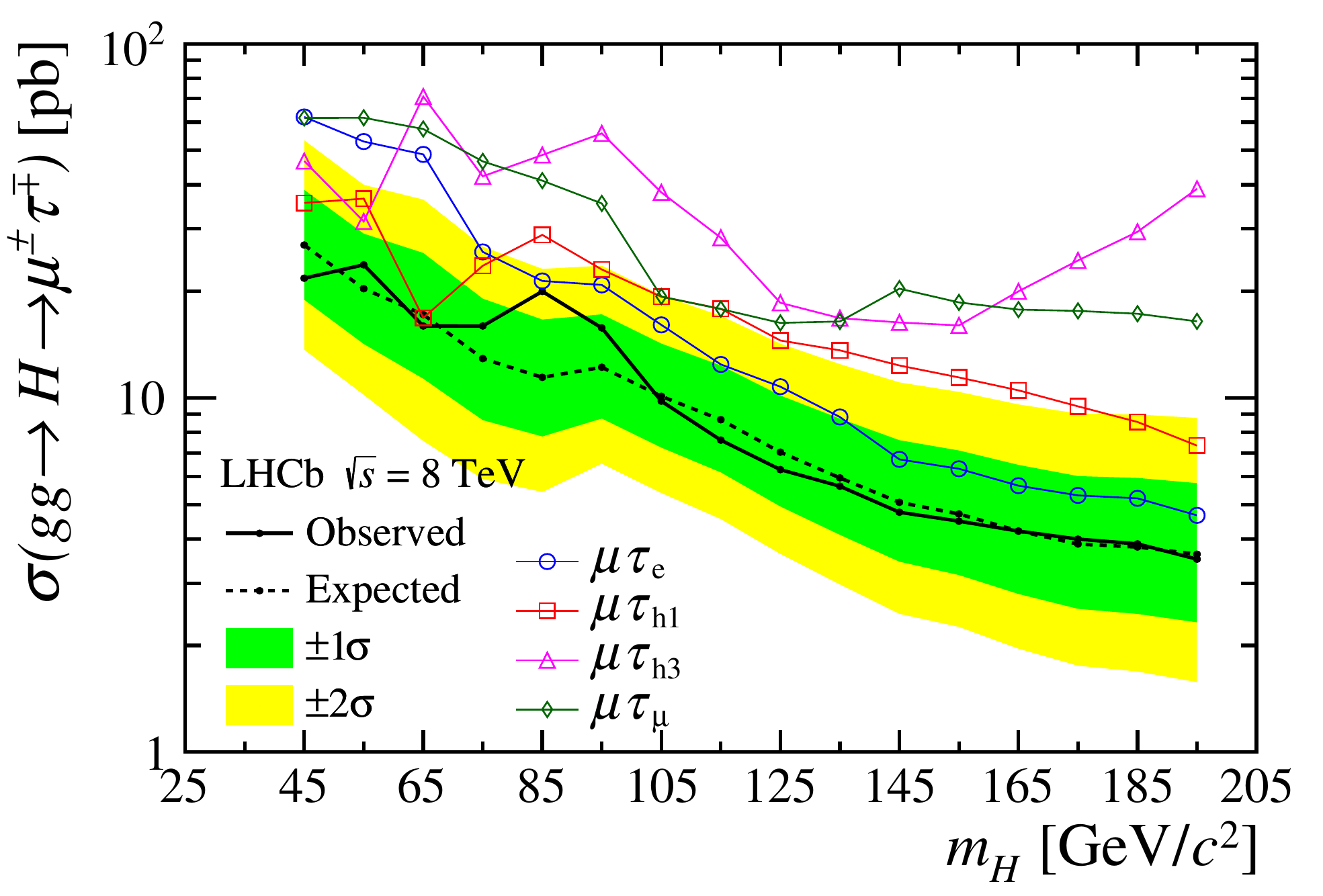}
\caption{\label{fig:lim}
  Cross-section times branching fraction 95\% CL limits for the \HMT decay
  as a function of \mH, from the simultaneous fit.
  The observed limits from individual channels are also shown.
}
\end{figure}

The fit results for all \mH values are compatible with a null signal, 
hence cross-section upper limits are computed. 
The exclusion limits of \CscBR defined at 95\% confidence level are obtained 
from the \CLs method \cite{Read:2002hq}. 
As mentioned before, for each mass hypothesis the selection considered 
is that providing the smallest expected limit.
%
%
%
The \CscBR exclusion limits are shown in \cref{fig:lim},
ranging from 22\pb for \mbox{\mH = 45\gevcc} to 4\pb for \mbox{\mH = 195\gevcc}.
In the particular case of \mbox{\mH = 125\gevcc},
using the production cross-section from Ref.~\cite{Heinemeyer:2013tqa} 
gives a best fit for the branching fraction of
\mbox{$\BRHMT = -2^{+14}_{-12}\%$}
and an observed exclusion limit 
\mbox{$\BRHMT < 26\%$}.
The corresponding exclusion limit on the Yukawa coupling is
\mbox{$\sqrt{|Y_{\mu\tau}|^2 + |Y_{\tau\mu}|^2} < \num{1.7e-2}$},
assuming the decay width
\mbox{$\Gamma_\text{SM} = 4.1\mevcc$}~\cite{Denner:2011mq}.

\section{Conclusion}

A search for Higgs-like bosons decaying via 
a lepton-flavour-violating process \HMT in $pp$ collisions at \sqs = 8\tev
is presented, with the tau lepton reconstructed in leptonic and hadronic decay modes.
No signal has been found.
The upper bound on the cross-section multiplied by the branching fraction, 
at 95\% confidence level, ranges from 22\pb for a boson mass
of 45\gevcc, to 4\pb for 195\gevcc.
The search provides information complementary to the \atlas and \cms collaborations.

\section*{Acknowledgements}
%
%
\noindent We express our gratitude to our colleagues in the CERN
accelerator departments for the excellent performance of the LHC. We
thank the technical and administrative staff at the LHCb
institutes.
We acknowledge support from CERN and from the national agencies:
CAPES, CNPq, FAPERJ and FINEP (Brazil); 
MOST and NSFC (China); 
CNRS/IN2P3 (France); 
BMBF, DFG and MPG (Germany); 
INFN (Italy); 
NWO (Netherlands); 
MNiSW and NCN (Poland); 
MEN/IFA (Romania); 
MSHE (Russia); 
MinECo (Spain); 
SNSF and SER (Switzerland); 
NASU (Ukraine); 
STFC (United Kingdom); 
NSF (USA).
We acknowledge the computing resources that are provided by CERN, IN2P3
(France), KIT and DESY (Germany), INFN (Italy), SURF (Netherlands),
PIC (Spain), GridPP (United Kingdom), RRCKI and Yandex
LLC (Russia), CSCS (Switzerland), IFIN-HH (Romania), CBPF (Brazil),
PL-GRID (Poland) and OSC (USA).
We are indebted to the communities behind the multiple open-source
software packages on which we depend.
Individual groups or members have received support from
AvH Foundation (Germany);
EPLANET, Marie Sk\l{}odowska-Curie Actions and ERC (European Union);
ANR, Labex P2IO and OCEVU, and R\'{e}gion Auvergne-Rh\^{o}ne-Alpes (France);
Key Research Program of Frontier Sciences of CAS, CAS PIFI, and the Thousand Talents Program (China);
RFBR, RSF and Yandex LLC (Russia);
GVA, XuntaGal and GENCAT (Spain);
the Royal Society
and the Leverhulme Trust (United Kingdom);
Laboratory Directed Research and Development program of LANL (USA).

\addcontentsline{toc}{section}{References}
\setboolean{inbibliography}{true}
\bibliographystyle{LHCb}
\bibliography{main,manual,LHCb-PAPER,LHCb-CONF,LHCb-DP,LHCb-TDR}

\ifx\mcitethebibliography\mciteundefinedmacro
\PackageError{LHCb.bst}{mciteplus.sty has not been loaded}
{This bibstyle requires the use of the mciteplus package.}\fi
\providecommand{\href}[2]{#2}
\begin{mcitethebibliography}{10}
\mciteSetBstSublistMode{n}
\mciteSetBstMaxWidthForm{subitem}{\alph{mcitesubitemcount})}
\mciteSetBstSublistLabelBeginEnd{\mcitemaxwidthsubitemform\space}
{\relax}{\relax}

\bibitem{Blanke:2008zb}
M.~Blanke {\em et~al.}, \ifthenelse{\boolean{articletitles}}{\emph{{$\Delta
  F=2$ observables and fine-tuning in a warped extra dimension with custodial
  protection}}, }{}\href{https://doi.org/10.1088/1126-6708/2009/03/001}{JHEP
  \textbf{03} (2009) 001},
  \href{http://arxiv.org/abs/0809.1073}{{\normalfont\ttfamily
  arXiv:0809.1073}}\relax
\mciteBstWouldAddEndPuncttrue
\mciteSetBstMidEndSepPunct{\mcitedefaultmidpunct}
{\mcitedefaultendpunct}{\mcitedefaultseppunct}\relax
\EndOfBibitem
\bibitem{Giudice:2008uua}
G.~F. Giudice and O.~Lebedev,
  \ifthenelse{\boolean{articletitles}}{\emph{{Higgs-dependent Yukawa
  couplings}}, }{}\href{https://doi.org/10.1016/j.physletb.2008.05.062}{Phys.\
  Lett.\ B \textbf{665} (2008) 79},
  \href{http://arxiv.org/abs/0804.1753}{{\normalfont\ttfamily
  arXiv:0804.1753}}\relax
\mciteBstWouldAddEndPuncttrue
\mciteSetBstMidEndSepPunct{\mcitedefaultmidpunct}
{\mcitedefaultendpunct}{\mcitedefaultseppunct}\relax
\EndOfBibitem
\bibitem{AguilarSaavedra:2009mx}
J.~A. Aguilar-Saavedra, \ifthenelse{\boolean{articletitles}}{\emph{{A minimal
  set of top-Higgs anomalous couplings}},
  }{}\href{https://doi.org/10.1016/j.nuclphysb.2009.06.022}{Nucl.\ Phys.\ B
  \textbf{821} (2009) 215},
  \href{http://arxiv.org/abs/0904.2387}{{\normalfont\ttfamily
  arXiv:0904.2387}}\relax
\mciteBstWouldAddEndPuncttrue
\mciteSetBstMidEndSepPunct{\mcitedefaultmidpunct}
{\mcitedefaultendpunct}{\mcitedefaultseppunct}\relax
\EndOfBibitem
\bibitem{Albrecht:2009xr}
M.~E. Albrecht {\em et~al.},
  \ifthenelse{\boolean{articletitles}}{\emph{{Electroweak and flavour structure
  of a warped extra dimension with custodial protection}},
  }{}\href{https://doi.org/10.1088/1126-6708/2009/09/064}{JHEP \textbf{09}
  (2009) 064}, \href{http://arxiv.org/abs/0903.2415}{{\normalfont\ttfamily
  arXiv:0903.2415}}\relax
\mciteBstWouldAddEndPuncttrue
\mciteSetBstMidEndSepPunct{\mcitedefaultmidpunct}
{\mcitedefaultendpunct}{\mcitedefaultseppunct}\relax
\EndOfBibitem
\bibitem{Goudelis:2011un}
A.~Goudelis, O.~Lebedev, and J.-h. Park,
  \ifthenelse{\boolean{articletitles}}{\emph{{Higgs-induced lepton flavour
  violation}}, }{}\href{https://doi.org/10.1016/j.physletb.2011.12.059}{Phys.\
  Lett.\ B \textbf{707} (2012) 369},
  \href{http://arxiv.org/abs/1111.1715}{{\normalfont\ttfamily
  arXiv:1111.1715}}\relax
\mciteBstWouldAddEndPuncttrue
\mciteSetBstMidEndSepPunct{\mcitedefaultmidpunct}
{\mcitedefaultendpunct}{\mcitedefaultseppunct}\relax
\EndOfBibitem
\bibitem{McKeen:2012av}
D.~McKeen, M.~Pospelov, and A.~Ritz,
  \ifthenelse{\boolean{articletitles}}{\emph{{Modified Higgs branching ratios
  versus CP and lepton flavour violation}},
  }{}\href{https://doi.org/10.1103/PhysRevD.86.113004}{Phys.\ Rev.\ D
  \textbf{86} (2012) 113004},
  \href{http://arxiv.org/abs/1208.4597}{{\normalfont\ttfamily
  arXiv:1208.4597}}\relax
\mciteBstWouldAddEndPuncttrue
\mciteSetBstMidEndSepPunct{\mcitedefaultmidpunct}
{\mcitedefaultendpunct}{\mcitedefaultseppunct}\relax
\EndOfBibitem
\bibitem{Arganda:2004bz}
E.~Arganda, A.~M. Curiel, M.~J. Herrero, and D.~Temes,
  \ifthenelse{\boolean{articletitles}}{\emph{{Lepton flavour violating Higgs
  boson decays from massive seesaw neutrinos}},
  }{}\href{https://doi.org/10.1103/PhysRevD.71.035011}{Phys.\ Rev.\ D
  \textbf{71} (2005) 035011},
  \href{http://arxiv.org/abs/hep-ph/0407302}{{\normalfont\ttfamily
  arXiv:hep-ph/0407302}}\relax
\mciteBstWouldAddEndPuncttrue
\mciteSetBstMidEndSepPunct{\mcitedefaultmidpunct}
{\mcitedefaultendpunct}{\mcitedefaultseppunct}\relax
\EndOfBibitem
\bibitem{Arganda:2014dta}
E.~Arganda, M.~J. Herrero, X.~Marcano, and C.~Weiland,
  \ifthenelse{\boolean{articletitles}}{\emph{{Imprints of massive inverse
  seesaw model neutrinos in lepton flavour violating Higgs boson decays}},
  }{}\href{https://doi.org/10.1103/PhysRevD.91.015001}{Phys.\ Rev.\ D
  \textbf{91} (2015) 015001},
  \href{http://arxiv.org/abs/1405.4300}{{\normalfont\ttfamily
  arXiv:1405.4300}}\relax
\mciteBstWouldAddEndPuncttrue
\mciteSetBstMidEndSepPunct{\mcitedefaultmidpunct}
{\mcitedefaultendpunct}{\mcitedefaultseppunct}\relax
\EndOfBibitem
\bibitem{Harnik:2012pb}
R.~Harnik, J.~Kopp, and J.~Zupan,
  \ifthenelse{\boolean{articletitles}}{\emph{{Flavour violating Higgs decays}},
  }{}\href{https://doi.org/10.1007/JHEP03(2013)026}{JHEP \textbf{03} (2013)
  026}, \href{http://arxiv.org/abs/1209.1397}{{\normalfont\ttfamily
  arXiv:1209.1397}}\relax
\mciteBstWouldAddEndPuncttrue
\mciteSetBstMidEndSepPunct{\mcitedefaultmidpunct}
{\mcitedefaultendpunct}{\mcitedefaultseppunct}\relax
\EndOfBibitem
\bibitem{Bjorken:1977vt}
J.~D. Bjorken and S.~Weinberg, \ifthenelse{\boolean{articletitles}}{\emph{{A
  mechanism for nonconservation of muon number}},
  }{}\href{https://doi.org/10.1103/PhysRevLett.38.622}{Phys.\ Rev.\ Lett.\
  \textbf{38} (1977) 622}\relax
\mciteBstWouldAddEndPuncttrue
\mciteSetBstMidEndSepPunct{\mcitedefaultmidpunct}
{\mcitedefaultendpunct}{\mcitedefaultseppunct}\relax
\EndOfBibitem
\bibitem{DiazCruz:1999xe}
J.~L. Diaz-Cruz and J.~J. Toscano,
  \ifthenelse{\boolean{articletitles}}{\emph{{Lepton flavour violating decays
  of Higgs bosons beyond the standard model}},
  }{}\href{https://doi.org/10.1103/PhysRevD.62.116005}{Phys.\ Rev.\ D
  \textbf{62} (2000) 116005},
  \href{http://arxiv.org/abs/hep-ph/9910233}{{\normalfont\ttfamily
  arXiv:hep-ph/9910233}}\relax
\mciteBstWouldAddEndPuncttrue
\mciteSetBstMidEndSepPunct{\mcitedefaultmidpunct}
{\mcitedefaultendpunct}{\mcitedefaultseppunct}\relax
\EndOfBibitem
\bibitem{Han:2000jz}
T.~Han and D.~Marfatia,
  \ifthenelse{\boolean{articletitles}}{\emph{{$H\to\mu\tau$ at hadron
  colliders}}, }{}\href{https://doi.org/10.1103/PhysRevLett.86.1442}{Phys.\
  Rev.\ Lett.\  \textbf{86} (2001) 1442},
  \href{http://arxiv.org/abs/hep-ph/0008141}{{\normalfont\ttfamily
  arXiv:hep-ph/0008141}}\relax
\mciteBstWouldAddEndPuncttrue
\mciteSetBstMidEndSepPunct{\mcitedefaultmidpunct}
{\mcitedefaultendpunct}{\mcitedefaultseppunct}\relax
\EndOfBibitem
\bibitem{Arhrib:2012ax}
A.~Arhrib, Y.~Cheng, and O.~C.~W. Kong,
  \ifthenelse{\boolean{articletitles}}{\emph{{Comprehensive analysis on lepton
  flavour violating Higgs boson to $\mu^\mp \tau^\pm$ decay in supersymmetry
  without $R$ parity}},
  }{}\href{https://doi.org/10.1103/PhysRevD.87.015025}{Phys.\ Rev.\ D
  \textbf{87} (2013) 015025},
  \href{http://arxiv.org/abs/1210.8241}{{\normalfont\ttfamily
  arXiv:1210.8241}}\relax
\mciteBstWouldAddEndPuncttrue
\mciteSetBstMidEndSepPunct{\mcitedefaultmidpunct}
{\mcitedefaultendpunct}{\mcitedefaultseppunct}\relax
\EndOfBibitem
\bibitem{Arana-Catania:2013xma}
M.~Arana-Catania, E.~Arganda, and M.~J. Herrero,
  \ifthenelse{\boolean{articletitles}}{\emph{{Non-decoupling SUSY in LFV Higgs
  decays: a window to new physics at the LHC}},
  }{}\href{https://doi.org/10.1007/JHEP09(2013)160}{JHEP \textbf{09} (2013)
  160}, Erratum \href{https://doi.org/10.1007/JHEP10(2015)192}{ibid.\
  \textbf{10} (2015) 192},
  \href{http://arxiv.org/abs/1304.3371}{{\normalfont\ttfamily
  arXiv:1304.3371}}\relax
\mciteBstWouldAddEndPuncttrue
\mciteSetBstMidEndSepPunct{\mcitedefaultmidpunct}
{\mcitedefaultendpunct}{\mcitedefaultseppunct}\relax
\EndOfBibitem
\bibitem{Agashe:2009di}
K.~Agashe and R.~Contino, \ifthenelse{\boolean{articletitles}}{\emph{{Composite
  Higgs-mediated flavor-changing neutral current}},
  }{}\href{https://doi.org/10.1103/PhysRevD.80.075016}{Phys.\ Rev.\ D
  \textbf{80} (2009) 075016},
  \href{http://arxiv.org/abs/0906.1542}{{\normalfont\ttfamily
  arXiv:0906.1542}}\relax
\mciteBstWouldAddEndPuncttrue
\mciteSetBstMidEndSepPunct{\mcitedefaultmidpunct}
{\mcitedefaultendpunct}{\mcitedefaultseppunct}\relax
\EndOfBibitem
\bibitem{Azatov:2009na}
A.~Azatov, M.~Toharia, and L.~Zhu,
  \ifthenelse{\boolean{articletitles}}{\emph{{Higgs mediated flavor-changing
  neutral current's in warped extra dimensions}},
  }{}\href{https://doi.org/10.1103/PhysRevD.80.035016}{Phys.\ Rev.\ D
  \textbf{80} (2009) 035016},
  \href{http://arxiv.org/abs/0906.1990}{{\normalfont\ttfamily
  arXiv:0906.1990}}\relax
\mciteBstWouldAddEndPuncttrue
\mciteSetBstMidEndSepPunct{\mcitedefaultmidpunct}
{\mcitedefaultendpunct}{\mcitedefaultseppunct}\relax
\EndOfBibitem
\bibitem{Perez:2008ee}
G.~Perez and L.~Randall, \ifthenelse{\boolean{articletitles}}{\emph{{Natural
  neutrino masses and mixings from warped geometry}},
  }{}\href{https://doi.org/10.1088/1126-6708/2009/01/077}{JHEP \textbf{01}
  (2009) 077}, \href{http://arxiv.org/abs/0805.4652}{{\normalfont\ttfamily
  arXiv:0805.4652}}\relax
\mciteBstWouldAddEndPuncttrue
\mciteSetBstMidEndSepPunct{\mcitedefaultmidpunct}
{\mcitedefaultendpunct}{\mcitedefaultseppunct}\relax
\EndOfBibitem
\bibitem{Casagrande:2008hr}
S.~Casagrande {\em et~al.}, \ifthenelse{\boolean{articletitles}}{\emph{{Flavour
  physics in the Randall-Sundrum model: I. Theoretical setup and electroweak
  precision tests}},
  }{}\href{https://doi.org/10.1088/1126-6708/2008/10/094}{JHEP \textbf{10}
  (2008) 094}, \href{http://arxiv.org/abs/0807.4937}{{\normalfont\ttfamily
  arXiv:0807.4937}}\relax
\mciteBstWouldAddEndPuncttrue
\mciteSetBstMidEndSepPunct{\mcitedefaultmidpunct}
{\mcitedefaultendpunct}{\mcitedefaultseppunct}\relax
\EndOfBibitem
\bibitem{Ishimori:2010au}
H.~Ishimori {\em et~al.},
  \ifthenelse{\boolean{articletitles}}{\emph{{Non-Abelian discrete symmetries
  in particle physics}}, }{}\href{https://doi.org/10.1143/PTPS.183.1}{Prog.\
  Theor.\ Phys.\ Suppl.\  \textbf{183} (2010) 1},
  \href{http://arxiv.org/abs/1003.3552}{{\normalfont\ttfamily
  arXiv:1003.3552}}\relax
\mciteBstWouldAddEndPuncttrue
\mciteSetBstMidEndSepPunct{\mcitedefaultmidpunct}
{\mcitedefaultendpunct}{\mcitedefaultseppunct}\relax
\EndOfBibitem
\bibitem{ALEPHZ}
\aleph collaboration, D.~Decamp {\em et~al.},
  \ifthenelse{\boolean{articletitles}}{\emph{{Searches for new particles in $Z$
  decays using the ALEPH detector}},
  }{}\href{https://doi.org/10.1016/0370-1573(92)90177-2}{Phys.\ Rept.\
  \textbf{216} (1992) 253}\relax
\mciteBstWouldAddEndPuncttrue
\mciteSetBstMidEndSepPunct{\mcitedefaultmidpunct}
{\mcitedefaultendpunct}{\mcitedefaultseppunct}\relax
\EndOfBibitem
\bibitem{DELPHIZ}
\delphi collaboration, P.~Abreu {\em et~al.},
  \ifthenelse{\boolean{articletitles}}{\emph{{A search for lepton flavour
  violation in \Z decays}},
  }{}\href{https://doi.org/10.1016/0370-2693(93)91737-8}{Phys.\ Lett.\ B
  \textbf{298} (1993) 247}\relax
\mciteBstWouldAddEndPuncttrue
\mciteSetBstMidEndSepPunct{\mcitedefaultmidpunct}
{\mcitedefaultendpunct}{\mcitedefaultseppunct}\relax
\EndOfBibitem
\bibitem{L3Z}
L3 collaboration, O.~Adriani {\em et~al.},
  \ifthenelse{\boolean{articletitles}}{\emph{{Search for lepton flavour
  violation in \Z decays}},
  }{}\href{https://doi.org/10.1016/0370-2693(93)90348-L}{Phys.\ Lett.\ B
  \textbf{316} (1993) 427}\relax
\mciteBstWouldAddEndPuncttrue
\mciteSetBstMidEndSepPunct{\mcitedefaultmidpunct}
{\mcitedefaultendpunct}{\mcitedefaultseppunct}\relax
\EndOfBibitem
\bibitem{OPALZ}
\opal collaboration, M.~Z. Akrawy {\em et~al.},
  \ifthenelse{\boolean{articletitles}}{\emph{{A Search for lepton flavour
  violation in \Z decays}},
  }{}\href{https://doi.org/10.1016/0370-2693(91)90437-U}{Phys.\ Lett.\ B
  \textbf{254} (1991) 293}\relax
\mciteBstWouldAddEndPuncttrue
\mciteSetBstMidEndSepPunct{\mcitedefaultmidpunct}
{\mcitedefaultendpunct}{\mcitedefaultseppunct}\relax
\EndOfBibitem
\bibitem{Abbiendi:2001cs}
\opal collaboration, G.~Abbiendi {\em et~al.},
  \ifthenelse{\boolean{articletitles}}{\emph{{Search for lepton flavour
  violation in \epem collisions at \sqs = 189--209 GeV}},
  }{}\href{https://doi.org/10.1016/S0370-2693(01)01086-3}{Phys.\ Lett.\ B
  \textbf{519} (2001) 23},
  \href{http://arxiv.org/abs/hep-ex/0109011}{{\normalfont\ttfamily
  arXiv:hep-ex/0109011}}\relax
\mciteBstWouldAddEndPuncttrue
\mciteSetBstMidEndSepPunct{\mcitedefaultmidpunct}
{\mcitedefaultendpunct}{\mcitedefaultseppunct}\relax
\EndOfBibitem
\bibitem{Blankenburg:2012ex}
G.~Blankenburg, J.~Ellis, and G.~Isidori,
  \ifthenelse{\boolean{articletitles}}{\emph{{Flavour-changing decays of a 125
  GeV Higgs-like particle}},
  }{}\href{https://doi.org/10.1016/j.physletb.2012.05.007}{Phys.\ Lett.\ B
  \textbf{712} (2012) 386},
  \href{http://arxiv.org/abs/1202.5704}{{\normalfont\ttfamily
  arXiv:1202.5704}}\relax
\mciteBstWouldAddEndPuncttrue
\mciteSetBstMidEndSepPunct{\mcitedefaultmidpunct}
{\mcitedefaultendpunct}{\mcitedefaultseppunct}\relax
\EndOfBibitem
\bibitem{Sirunyan:2017xzt}
CMS collaboration, A.~M. Sirunyan {\em et~al.},
  \ifthenelse{\boolean{articletitles}}{\emph{{Search for lepton flavour
  violating decays of the Higgs boson to $\mu\tau$ and e$\tau$ in proton-proton
  collisions at $\sqrt{s}=$ 13 TeV}},
  }{}\href{https://doi.org/10.1007/JHEP06(2018)001}{JHEP \textbf{06} (2018)
  001}, \href{http://arxiv.org/abs/1712.07173}{{\normalfont\ttfamily
  arXiv:1712.07173}}\relax
\mciteBstWouldAddEndPuncttrue
\mciteSetBstMidEndSepPunct{\mcitedefaultmidpunct}
{\mcitedefaultendpunct}{\mcitedefaultseppunct}\relax
\EndOfBibitem
\bibitem{Aad:2015gha}
ATLAS collaboration, G.~Aad {\em et~al.},
  \ifthenelse{\boolean{articletitles}}{\emph{{Search for
  lepton-flavour-violating $H\to\muon\tauon$ decays of the Higgs boson with the
  ATLAS detector}}, }{}\href{https://doi.org/10.1007/JHEP11(2015)211}{JHEP
  \textbf{11} (2015) 211},
  \href{http://arxiv.org/abs/1508.03372}{{\normalfont\ttfamily
  arXiv:1508.03372}}\relax
\mciteBstWouldAddEndPuncttrue
\mciteSetBstMidEndSepPunct{\mcitedefaultmidpunct}
{\mcitedefaultendpunct}{\mcitedefaultseppunct}\relax
\EndOfBibitem
\bibitem{GUNION2HDM}
J.~F. Gunion and H.~E. Haber, \ifthenelse{\boolean{articletitles}}{\emph{{The
  CP-conserving two-Higgs doublet model: The Approach to the decoupling
  limit}}, }{}\href{https://doi.org/10.1103/PhysRevD.67.075019}{Phys.\ Rev.\ D
  \textbf{67} (2003) 075019},
  \href{http://arxiv.org/abs/hep-ph/0207010}{{\normalfont\ttfamily
  arXiv:hep-ph/0207010}}\relax
\mciteBstWouldAddEndPuncttrue
\mciteSetBstMidEndSepPunct{\mcitedefaultmidpunct}
{\mcitedefaultendpunct}{\mcitedefaultseppunct}\relax
\EndOfBibitem
\bibitem{Aad:2012cfr}
ATLAS collaboration, G.~Aad {\em et~al.},
  \ifthenelse{\boolean{articletitles}}{\emph{{Search for the neutral Higgs
  bosons of the Minimal Supersymmetric Standard Model in $pp$ collisions at
  $\sqrt{s}=7$ TeV with the ATLAS detector}},
  }{}\href{https://doi.org/10.1007/JHEP02(2013)095}{JHEP \textbf{02} (2013)
  095}, \href{http://arxiv.org/abs/1211.6956}{{\normalfont\ttfamily
  arXiv:1211.6956}}\relax
\mciteBstWouldAddEndPuncttrue
\mciteSetBstMidEndSepPunct{\mcitedefaultmidpunct}
{\mcitedefaultendpunct}{\mcitedefaultseppunct}\relax
\EndOfBibitem
\bibitem{Chatrchyan:2012vp}
CMS collaboration, S.~Chatrchyan {\em et~al.},
  \ifthenelse{\boolean{articletitles}}{\emph{{Search for neutral Higgs bosons
  decaying to tau pairs in $pp$ collisions at $\sqrt{s}=7$ TeV}},
  }{}\href{https://doi.org/10.1016/j.physletb.2012.05.028}{Phys.\ Lett.\ B
  \textbf{713} (2012) 68},
  \href{http://arxiv.org/abs/1202.4083}{{\normalfont\ttfamily
  arXiv:1202.4083}}\relax
\mciteBstWouldAddEndPuncttrue
\mciteSetBstMidEndSepPunct{\mcitedefaultmidpunct}
{\mcitedefaultendpunct}{\mcitedefaultseppunct}\relax
\EndOfBibitem
\bibitem{Jaeckel:2010ni}
J.~Jaeckel and A.~Ringwald, \ifthenelse{\boolean{articletitles}}{\emph{{The
  low-energy frontier of particle physics}},
  }{}\href{https://doi.org/10.1146/annurev.nucl.012809.104433}{Ann.\ Rev.\
  Nucl.\ Part.\ Sci.\  \textbf{60} (2010) 405},
  \href{http://arxiv.org/abs/1002.0329}{{\normalfont\ttfamily
  arXiv:1002.0329}}\relax
\mciteBstWouldAddEndPuncttrue
\mciteSetBstMidEndSepPunct{\mcitedefaultmidpunct}
{\mcitedefaultendpunct}{\mcitedefaultseppunct}\relax
\EndOfBibitem
\bibitem{Baumgart:2009tn}
M.~Baumgart {\em et~al.},
  \ifthenelse{\boolean{articletitles}}{\emph{{Non-Abelian dark sectors and
  their collider signatures}},
  }{}\href{https://doi.org/10.1088/1126-6708/2009/04/014}{JHEP \textbf{04}
  (2009) 014}, \href{http://arxiv.org/abs/0901.0283}{{\normalfont\ttfamily
  arXiv:0901.0283}}\relax
\mciteBstWouldAddEndPuncttrue
\mciteSetBstMidEndSepPunct{\mcitedefaultmidpunct}
{\mcitedefaultendpunct}{\mcitedefaultseppunct}\relax
\EndOfBibitem
\bibitem{BABARDark}
\babar collaboration, J.~P. Lees {\em et~al.},
  \ifthenelse{\boolean{articletitles}}{\emph{{Search for low-mass dark-sector
  Higgs bosons}},
  }{}\href{https://doi.org/10.1103/PhysRevLett.108.211801}{Phys.\ Rev.\ Lett.\
  \textbf{108} (2012) 211801},
  \href{http://arxiv.org/abs/1202.1313}{{\normalfont\ttfamily
  arXiv:1202.1313}}\relax
\mciteBstWouldAddEndPuncttrue
\mciteSetBstMidEndSepPunct{\mcitedefaultmidpunct}
{\mcitedefaultendpunct}{\mcitedefaultseppunct}\relax
\EndOfBibitem
\bibitem{BELLEDark}
\belle collaboration, I.~Jaegle,
  \ifthenelse{\boolean{articletitles}}{\emph{{Search for the 'Dark Photon' and
  the 'Dark Higgs' at Belle}},
  }{}\href{https://doi.org/10.1016/j.nuclphysbps.2012.11.008}{Nucl.\ Phys.\
  Proc.\ Suppl.\  \textbf{234} (2013) 33},
  \href{http://arxiv.org/abs/1211.1403}{{\normalfont\ttfamily
  arXiv:1211.1403}}\relax
\mciteBstWouldAddEndPuncttrue
\mciteSetBstMidEndSepPunct{\mcitedefaultmidpunct}
{\mcitedefaultendpunct}{\mcitedefaultseppunct}\relax
\EndOfBibitem
\bibitem{LHCb-PAPER-2017-038}
LHCb collaboration, R.~Aaij {\em et~al.},
  \ifthenelse{\boolean{articletitles}}{\emph{{Search for dark photons produced
  in 13\tev $pp$ collisions}},
  }{}\href{https://doi.org/10.1103/PhysRevLett.120.061801}{Phys.\ Rev.\ Lett.\
  \textbf{120} (2018) 061801},
  \href{http://arxiv.org/abs/1710.02867}{{\normalfont\ttfamily
  arXiv:1710.02867}}\relax
\mciteBstWouldAddEndPuncttrue
\mciteSetBstMidEndSepPunct{\mcitedefaultmidpunct}
{\mcitedefaultendpunct}{\mcitedefaultseppunct}\relax
\EndOfBibitem
\bibitem{Georgi:1977gs}
H.~M. Georgi, S.~L. Glashow, M.~E. Machacek, and D.~V. Nanopoulos,
  \ifthenelse{\boolean{articletitles}}{\emph{{Higgs bosons from two-gluon
  annihilation in proton-proton collisions}},
  }{}\href{https://doi.org/10.1103/PhysRevLett.40.692}{Phys.\ Rev.\ Lett.\
  \textbf{40} (1978) 692}\relax
\mciteBstWouldAddEndPuncttrue
\mciteSetBstMidEndSepPunct{\mcitedefaultmidpunct}
{\mcitedefaultendpunct}{\mcitedefaultseppunct}\relax
\EndOfBibitem
\bibitem{LHCb-PAPER-2018-016}
LHCb collaboration, R.~Aaij {\em et~al.},
  \ifthenelse{\boolean{articletitles}}{\emph{{Measurement of
  \decay{Z}{\taup\taum} production in proton-proton collisions at $\sqrt{s} =
  8$\tev}}, }{}\href{https://doi.org/10.1007/JHEP09(2018)159}{JHEP \textbf{09}
  (2018) 159}, \href{http://arxiv.org/abs/1806.05008}{{\normalfont\ttfamily
  arXiv:1806.05008}}\relax
\mciteBstWouldAddEndPuncttrue
\mciteSetBstMidEndSepPunct{\mcitedefaultmidpunct}
{\mcitedefaultendpunct}{\mcitedefaultseppunct}\relax
\EndOfBibitem
\bibitem{Alves:2008zz}
LHCb collaboration, A.~A. Alves~Jr.\ {\em et~al.},
  \ifthenelse{\boolean{articletitles}}{\emph{{The \lhcb detector at the LHC}},
  }{}\href{https://doi.org/10.1088/1748-0221/3/08/S08005}{JINST \textbf{3}
  (2008) S08005}\relax
\mciteBstWouldAddEndPuncttrue
\mciteSetBstMidEndSepPunct{\mcitedefaultmidpunct}
{\mcitedefaultendpunct}{\mcitedefaultseppunct}\relax
\EndOfBibitem
\bibitem{LHCb-DP-2014-002}
LHCb collaboration, R.~Aaij {\em et~al.},
  \ifthenelse{\boolean{articletitles}}{\emph{{LHCb detector performance}},
  }{}\href{https://doi.org/10.1142/S0217751X15300227}{Int.\ J.\ Mod.\ Phys.\ A
  \textbf{30} (2015) 1530022},
  \href{http://arxiv.org/abs/1412.6352}{{\normalfont\ttfamily
  arXiv:1412.6352}}\relax
\mciteBstWouldAddEndPuncttrue
\mciteSetBstMidEndSepPunct{\mcitedefaultmidpunct}
{\mcitedefaultendpunct}{\mcitedefaultseppunct}\relax
\EndOfBibitem
\bibitem{Sjostrand:2007gs}
T.~Sj\"{o}strand, S.~Mrenna, and P.~Skands,
  \ifthenelse{\boolean{articletitles}}{\emph{{A brief introduction to PYTHIA
  8.1}}, }{}\href{https://doi.org/10.1016/j.cpc.2008.01.036}{Comput.\ Phys.\
  Commun.\  \textbf{178} (2008) 852},
  \href{http://arxiv.org/abs/0710.3820}{{\normalfont\ttfamily
  arXiv:0710.3820}}\relax
\mciteBstWouldAddEndPuncttrue
\mciteSetBstMidEndSepPunct{\mcitedefaultmidpunct}
{\mcitedefaultendpunct}{\mcitedefaultseppunct}\relax
\EndOfBibitem
\bibitem{Sjostrand:2006za}
T.~Sj\"{o}strand, S.~Mrenna, and P.~Skands,
  \ifthenelse{\boolean{articletitles}}{\emph{{PYTHIA 6.4 physics and manual}},
  }{}\href{https://doi.org/10.1088/1126-6708/2006/05/026}{JHEP \textbf{05}
  (2006) 026}, \href{http://arxiv.org/abs/hep-ph/0603175}{{\normalfont\ttfamily
  arXiv:hep-ph/0603175}}\relax
\mciteBstWouldAddEndPuncttrue
\mciteSetBstMidEndSepPunct{\mcitedefaultmidpunct}
{\mcitedefaultendpunct}{\mcitedefaultseppunct}\relax
\EndOfBibitem
\bibitem{LHCb-PROC-2010-056}
I.~Belyaev {\em et~al.}, \ifthenelse{\boolean{articletitles}}{\emph{{Handling
  of the generation of primary events in Gauss, the LHCb simulation
  framework}}, }{}\href{https://doi.org/10.1088/1742-6596/331/3/032047}{J.\
  Phys.\ Conf.\ Ser.\  \textbf{331} (2011) 032047}\relax
\mciteBstWouldAddEndPuncttrue
\mciteSetBstMidEndSepPunct{\mcitedefaultmidpunct}
{\mcitedefaultendpunct}{\mcitedefaultseppunct}\relax
\EndOfBibitem
\bibitem{cteq6l}
J.~Pumplin {\em et~al.}, \ifthenelse{\boolean{articletitles}}{\emph{{New
  generation of parton distributions with uncertainties from global QCD
  analysis}}, }{}\href{https://doi.org/10.1088/1126-6708/2002/07/012}{JHEP
  \textbf{07} (2002) 012},
  \href{http://arxiv.org/abs/hep-ph/0201195}{{\normalfont\ttfamily
  arXiv:hep-ph/0201195}}\relax
\mciteBstWouldAddEndPuncttrue
\mciteSetBstMidEndSepPunct{\mcitedefaultmidpunct}
{\mcitedefaultendpunct}{\mcitedefaultseppunct}\relax
\EndOfBibitem
\bibitem{Lange:2001uf}
D.~J. Lange, \ifthenelse{\boolean{articletitles}}{\emph{{The EvtGen particle
  decay simulation package}},
  }{}\href{https://doi.org/10.1016/S0168-9002(01)00089-4}{Nucl.\ Instrum.\
  Meth.\ A \textbf{462} (2001) 152}\relax
\mciteBstWouldAddEndPuncttrue
\mciteSetBstMidEndSepPunct{\mcitedefaultmidpunct}
{\mcitedefaultendpunct}{\mcitedefaultseppunct}\relax
\EndOfBibitem
\bibitem{Golonka:2005pn}
P.~Golonka and Z.~Was, \ifthenelse{\boolean{articletitles}}{\emph{{PHOTOS Monte
  Carlo: A precision tool for QED corrections in $Z$ and $W$ decays}},
  }{}\href{https://doi.org/10.1140/epjc/s2005-02396-4}{Eur.\ Phys.\ J.\ C
  \textbf{45} (2006) 97},
  \href{http://arxiv.org/abs/hep-ph/0506026}{{\normalfont\ttfamily
  arXiv:hep-ph/0506026}}\relax
\mciteBstWouldAddEndPuncttrue
\mciteSetBstMidEndSepPunct{\mcitedefaultmidpunct}
{\mcitedefaultendpunct}{\mcitedefaultseppunct}\relax
\EndOfBibitem
\bibitem{Allison:2006ve}
Geant4 collaboration, J.~Allison {\em et~al.},
  \ifthenelse{\boolean{articletitles}}{\emph{{Geant4 developments and
  applications}}, }{}\href{https://doi.org/10.1109/TNS.2006.869826}{IEEE
  Trans.\ Nucl.\ Sci.\  \textbf{53} (2006) 270}\relax
\mciteBstWouldAddEndPuncttrue
\mciteSetBstMidEndSepPunct{\mcitedefaultmidpunct}
{\mcitedefaultendpunct}{\mcitedefaultseppunct}\relax
\EndOfBibitem
\bibitem{Agostinelli:2002hh}
Geant4 collaboration, S.~Agostinelli {\em et~al.},
  \ifthenelse{\boolean{articletitles}}{\emph{{Geant4: A simulation toolkit}},
  }{}\href{https://doi.org/10.1016/S0168-9002(03)01368-8}{Nucl.\ Instrum.\
  Meth.\ A \textbf{506} (2003) 250}\relax
\mciteBstWouldAddEndPuncttrue
\mciteSetBstMidEndSepPunct{\mcitedefaultmidpunct}
{\mcitedefaultendpunct}{\mcitedefaultseppunct}\relax
\EndOfBibitem
\bibitem{LHCb-PROC-2011-006}
M.~Clemencic {\em et~al.}, \ifthenelse{\boolean{articletitles}}{\emph{{The
  \lhcb simulation application, Gauss: Design, evolution and experience}},
  }{}\href{https://doi.org/10.1088/1742-6596/331/3/032023}{J.\ Phys.\ Conf.\
  Ser.\  \textbf{331} (2011) 032023}\relax
\mciteBstWouldAddEndPuncttrue
\mciteSetBstMidEndSepPunct{\mcitedefaultmidpunct}
{\mcitedefaultendpunct}{\mcitedefaultseppunct}\relax
\EndOfBibitem
\bibitem{inspire:659055}
P.~Nason, \ifthenelse{\boolean{articletitles}}{\emph{{A new method for
  combining NLO QCD with shower Monte Carlo algorithms}},
  }{}\href{https://doi.org/10.1088/1126-6708/2004/11/040}{JHEP \textbf{11}
  (2004) 040}, \href{http://arxiv.org/abs/hep-ph/0409146}{{\normalfont\ttfamily
  arXiv:hep-ph/0409146}}\relax
\mciteBstWouldAddEndPuncttrue
\mciteSetBstMidEndSepPunct{\mcitedefaultmidpunct}
{\mcitedefaultendpunct}{\mcitedefaultseppunct}\relax
\EndOfBibitem
\bibitem{inspire:760769}
S.~Frixione, P.~Nason, and C.~Oleari,
  \ifthenelse{\boolean{articletitles}}{\emph{{Matching NLO QCD computations
  with parton shower simulations: the POWHEG method}},
  }{}\href{https://doi.org/10.1088/1126-6708/2007/11/070}{JHEP \textbf{11}
  (2007) 070}, \href{http://arxiv.org/abs/0709.2092}{{\normalfont\ttfamily
  arXiv:0709.2092}}\relax
\mciteBstWouldAddEndPuncttrue
\mciteSetBstMidEndSepPunct{\mcitedefaultmidpunct}
{\mcitedefaultendpunct}{\mcitedefaultseppunct}\relax
\EndOfBibitem
\bibitem{inspire:845712}
S.~Alioli, P.~Nason, C.~Oleari, and E.~Re,
  \ifthenelse{\boolean{articletitles}}{\emph{{A general framework for
  implementing NLO calculations in shower Monte Carlo programs: the POWHEG
  BOX}}, }{}\href{https://doi.org/10.1007/JHEP06(2010)043}{JHEP \textbf{06}
  (2010) 043}, \href{http://arxiv.org/abs/1002.2581}{{\normalfont\ttfamily
  arXiv:1002.2581}}\relax
\mciteBstWouldAddEndPuncttrue
\mciteSetBstMidEndSepPunct{\mcitedefaultmidpunct}
{\mcitedefaultendpunct}{\mcitedefaultseppunct}\relax
\EndOfBibitem
\bibitem{inspire:804159}
S.~Alioli, P.~Nason, C.~Oleari, and E.~Re,
  \ifthenelse{\boolean{articletitles}}{\emph{{NLO Higgs boson production via
  gluon fusion matched with shower in POWHEG}},
  }{}\href{https://doi.org/10.1088/1126-6708/2009/04/002}{JHEP \textbf{04}
  (2009) 002}, \href{http://arxiv.org/abs/0812.0578}{{\normalfont\ttfamily
  arXiv:0812.0578}}\relax
\mciteBstWouldAddEndPuncttrue
\mciteSetBstMidEndSepPunct{\mcitedefaultmidpunct}
{\mcitedefaultendpunct}{\mcitedefaultseppunct}\relax
\EndOfBibitem
\bibitem{inspire:1334137}
L.~A. Harland-Lang, A.~D. Martin, P.~Motylinski, and R.~S. Thorne,
  \ifthenelse{\boolean{articletitles}}{\emph{{Parton distributions in the LHC
  era: MMHT 2014 PDFs}},
  }{}\href{https://doi.org/10.1140/epjc/s10052-015-3397-6}{Eur.\ Phys.\ J.\ C
  \textbf{75} (2015) 204},
  \href{http://arxiv.org/abs/1412.3989}{{\normalfont\ttfamily
  arXiv:1412.3989}}\relax
\mciteBstWouldAddEndPuncttrue
\mciteSetBstMidEndSepPunct{\mcitedefaultmidpunct}
{\mcitedefaultendpunct}{\mcitedefaultseppunct}\relax
\EndOfBibitem
\bibitem{LHCb-PAPER-2014-047}
LHCb collaboration, R.~Aaij {\em et~al.},
  \ifthenelse{\boolean{articletitles}}{\emph{{Precision luminosity measurements
  at LHCb}}, }{}\href{https://doi.org/10.1088/1748-0221/9/12/P12005}{JINST
  \textbf{9} (2014) P12005},
  \href{http://arxiv.org/abs/1410.0149}{{\normalfont\ttfamily
  arXiv:1410.0149}}\relax
\mciteBstWouldAddEndPuncttrue
\mciteSetBstMidEndSepPunct{\mcitedefaultmidpunct}
{\mcitedefaultendpunct}{\mcitedefaultseppunct}\relax
\EndOfBibitem
\bibitem{brem}
F.~Machefert, \ifthenelse{\boolean{articletitles}}{\emph{{LHCb} calorimeters
  and muon system lepton identification}, }{} in {\em {AIP} Conference
  Proceedings}, {AIP}, 2004.
\newblock
  doi:~\href{https://doi.org/10.1063/1.1807309}{10.1063/1.1807309}\relax
\mciteBstWouldAddEndPuncttrue
\mciteSetBstMidEndSepPunct{\mcitedefaultmidpunct}
{\mcitedefaultendpunct}{\mcitedefaultseppunct}\relax
\EndOfBibitem
\bibitem{Read:2002hq}
A.~L. Read, \ifthenelse{\boolean{articletitles}}{\emph{{Presentation of search
  results: The \CLs technique}},
  }{}\href{https://doi.org/10.1088/0954-3899/28/10/313}{J.\ Phys.\ G
  \textbf{28} (2002) 2693}\relax
\mciteBstWouldAddEndPuncttrue
\mciteSetBstMidEndSepPunct{\mcitedefaultmidpunct}
{\mcitedefaultendpunct}{\mcitedefaultseppunct}\relax
\EndOfBibitem
\bibitem{Heinemeyer:2013tqa}
LHC Higgs Cross Section Working Group, J.~R. Andersen {\em et~al.},
  \ifthenelse{\boolean{articletitles}}{\emph{{Handbook of LHC Higgs cross
  sections: 3. Higgs properties}}, }{} tech. rep., CERN, Jul, 2013.
\newblock doi:~\href{https://doi.org/CERN-2013-004}{CERN-2013-004}\relax
\mciteBstWouldAddEndPuncttrue
\mciteSetBstMidEndSepPunct{\mcitedefaultmidpunct}
{\mcitedefaultendpunct}{\mcitedefaultseppunct}\relax
\EndOfBibitem
\bibitem{Denner:2011mq}
A.~Denner {\em et~al.}, \ifthenelse{\boolean{articletitles}}{\emph{{Standard
  Model Higgs-boson branching ratios with uncertainties}},
  }{}\href{https://doi.org/10.1140/epjc/s10052-011-1753-8}{Eur.\ Phys.\ J.\ C
  \textbf{71} (2011) 1753},
  \href{http://arxiv.org/abs/1107.5909}{{\normalfont\ttfamily
  arXiv:1107.5909}}\relax
\mciteBstWouldAddEndPuncttrue
\mciteSetBstMidEndSepPunct{\mcitedefaultmidpunct}
{\mcitedefaultendpunct}{\mcitedefaultseppunct}\relax
\EndOfBibitem
\end{mcitethebibliography}

\newpage


\newpage
\centerline{\large\bf LHCb collaboration}
\begin{flushleft}
\small
R.~Aaij$^{27}$,
C.~Abell{\'a}n~Beteta$^{44}$,
B.~Adeva$^{41}$,
M.~Adinolfi$^{48}$,
C.A.~Aidala$^{73}$,
Z.~Ajaltouni$^{5}$,
S.~Akar$^{59}$,
P.~Albicocco$^{18}$,
J.~Albrecht$^{10}$,
F.~Alessio$^{42}$
% ...

{\footnotesize \it
% ...
$ ^{5}$Clermont Universit{\'e}, Universit{\'e} Blaise Pascal, CNRS/IN2P3, LPC, Clermont-Ferrand, France\\
% ...
$ ^{10}$Fakult{\"a}t Physik, Technische Universit{\"a}t Dortmund, Dortmund, Germany\\
% ...
$ ^{18}$INFN Laboratori Nazionali di Frascati, Frascati, Italy\\
% ...
$ ^{27}$Nikhef National Institute for Subatomic Physics, Amsterdam, Netherlands\\
% ...
$ ^{41}$Instituto Galego de F{\'\i}sica de Altas Enerx{\'\i}as (IGFAE), Universidade de Santiago de Compostela, Santiago de Compostela, Spain\\
$ ^{42}$European Organization for Nuclear Research (CERN), Geneva, Switzerland\\
% ...
$ ^{44}$Physik-Institut, Universit{\"a}t Z{\"u}rich, Z{\"u}rich, Switzerland\\
% ...
$ ^{48}$H.H. Wills Physics Laboratory, University of Bristol, Bristol, United Kingdom\\
% ...
$ ^{59}$University of Cincinnati, Cincinnati, OH, United States\\
% ...
$ ^{61}$Syracuse University, Syracuse, NY, United States\\
% ...
$ ^{73}$University of Michigan, Ann Arbor, United States, associated to $^{61}$\\
% ...
}
\end{flushleft}

\end{document}